%% file: main.tex
\documentclass[sigconf]{acmart}

\copyrightyear{2025}
\acmYear{2025}
\setcopyright{cc}
\setcctype{by-nc-sa}
\acmConference[MICRO'25]{58th IEEE/ACM International Symposium on Microarchitecture}{October 18-22, 2025}{Seoul, Republic of Korea}
\acmBooktitle{58th IEEE/ACM International Symposium on Microarchitecture (MICRO'25), October 18-22, 2025, Seoul, Republic of Korea}\acmDOI{10.1145/3725843.3756064}
\acmISBN{979-8-4007-1573-0/2025/10}

\usepackage{enumitem}
\usepackage{textcomp}
\usepackage{subcaption}
\usepackage{multirow}
\usepackage{soul}
\usepackage{bm}
\usepackage[ruled,linesnumbered]{algorithm2e} 
\usepackage{nicefrac} 
\usepackage{cleveref}

\usepackage{listings}
\usepackage{pifont}

\def\pname{\textsc{Elk}\xspace}
\def\naive{{\textit{Basic}}\xspace}
\def\baseline{{\textit{Static}}\xspace}
\def\fixorder{{\textit{\pname-Dyn}}\xspace}
\def\full{{\textit{\pname-Full}}\xspace}
\def\ideal{{\textit{Ideal}}\xspace}
\newcommand{\textstt}[1]{\begin{small}\texttt{#1}\end{small}}

\DeclareRobustCommand{\hlcommon}[1]{\sethlcolor{white}\hl{#1}}
\DeclareRobustCommand{\hlA}[1]{\sethlcolor{white}\hl{#1}}
\DeclareRobustCommand{\hlB}[1]{\sethlcolor{white}\hl{#1}}
\DeclareRobustCommand{\hlC}[1]{\sethlcolor{white}\hl{#1}}
\DeclareRobustCommand{\hlD}[1]{\sethlcolor{white}\hl{#1}}
\DeclareRobustCommand{\hlE}[1]{\sethlcolor{white}\hl{#1}}
\DeclareRobustCommand{\hlF}[1]{\sethlcolor{white}\hl{#1}}

\let\oldding\ding
\renewcommand{\ding}[2][1]{\scalebox{#1.1}{\oldding{#2}}}

\newcommand{\specialcell}[2][c]{%
  \begin{tabular}[#1]{@{}c@{}}#2\end{tabular}}



\title{\pname{}: Exploring the Efficiency of Inter-core Connected AI Chips with Deep Learning Compiler Techniques}

\input{author}

\begin{document}




\input{abstract}

\maketitle


\input{intro}

\input{background}
\input{motivation}

\input{design}
\input{implt}

\input{eval}
\input{discussion}
\input{related}
\input{conclusion}


\newpage

\balance
\bibliographystyle{ACM-Reference-Format}
\bibliography{refs}

\input{appendix}

\end{document}

%% file: author.tex
\author{Yiqi Liu}
\email{yiqiliu2@illinois.edu}
\affiliation{
  \institution{UIUC}
  \city{}
  \state{}
  \country{}
}

\author{Yuqi Xue}
\email{yuqixue2@illinois.edu}
\affiliation{
  \institution{UIUC}
  \city{}
  \state{}
  \country{}
}

\author{Noelle Crawford}
\email{noellec3@illinois.edu}
\affiliation{
  \institution{UIUC}
  \city{}
  \state{}
  \country{}
}

\author{Jilong Xue}
\email{jxue@microsoft.com}
\affiliation{
  \institution{Microsoft Research}
  \city{}
  \state{}
  \country{}
}

\author{Jian Huang}
\email{jianh@illinois.edu}
\affiliation{
  \institution{UIUC}
  \city{}
  \state{}
  \country{}
}

%% file: abstract.tex
\begin{abstract}
\vspace{-0.5ex}

To meet the increasing demand of deep learning (DL) models, AI chips are employing both off-chip memory (e.g., HBM) and high-bandwidth low-latency interconnect for direct inter-core data exchange. 
However, it is not easy to explore the efficiency of these \underline{\textbf{i}}nter-\underline{\textbf{c}}ore \underline{\textbf{c}}onnected \underline{\textbf{A}}I (ICCA) chips, 
 due to a fundamental tussle among compute (per-core execution), communication (inter-core data exchange), and I/O (off-chip data access). 

In this paper, we develop \pname{}, a DL compiler framework to maximize the efficiency of ICCA chips by jointly trading off all the three performance factors discussed above. 
\pname{} structures these performance factors into configurable parameters and forms a global trade-off space in the DL compiler.
To systematically explore this space and maximize overall efficiency, \pname{} employs a new inductive operator scheduling policy and a cost-aware on-chip memory allocation algorithm. It generates globally optimized execution plans that best overlap off-chip data loading and on-chip execution.
To examine the efficiency of \pname{}, we build a full-fledged emulator based on a real ICCA chip IPU-POD4, and an ICCA chip simulator for sensitivity analysis with different interconnect network topologies.  
\pname{} achieves 94\% of the ideal roofline performance of ICCA chips on average, 
showing the benefits of supporting large DL models on ICCA chips. We also show \pname{}'s capability of enabling architecture design space exploration for new ICCA chip development.  

\end{abstract}

%% file: intro.tex
\vspace{-0.2em}
\section{Introduction}
\label{sec:intro}
\vspace{-0.1em}

To meet the ever-increasing compute demand of deep learning (DL) like large language models (LLMs)~\cite{megatron,vllm}, various AI chips have been developed~\cite{h100-hbm,ipu2,samba-sn40,meta-mtia,tenstorrent}.
A typical AI chip employs many parallel cores to scale computing throughput. Each core has its local SRAM as a scratchpad memory.
\hlF{To exploit this parallelism, the DL compiler partitions a tensor operator (e.g., BatchMatMul in attention and MatMul in FFN \mbox{\cite{attentionisallyouneed}}) into tiles and maps each tile to a core.}
\hlF{Since the on-chip SRAM size is limited, AI chips can employ off-chip memories (e.g., HBM) to provide larger capacity and accommodate the model parameters of larger DL models.}

\begin{figure}[t]
    \centering
    \vspace{2ex}
    \includegraphics[scale=0.8]{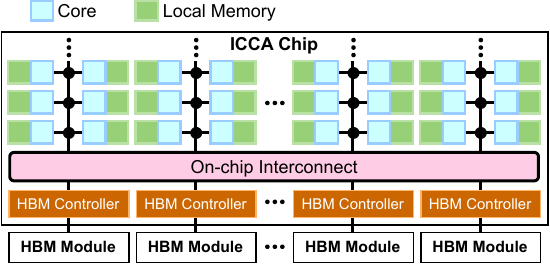} 
    \vspace{-1ex}
    \caption{Architecture of inter-core connected AI (ICCA) chip.}
    \label{fig:ipu-arch}
\end{figure}

However, the off-chip memory bandwidth scales much slower than compute performance, and cannot meet the growing demand of large models.
To alleviate the bandwidth bottleneck, inter-core connected AI (ICCA) chips were proposed. They enable inter-core links that allow one core to directly access data from other cores' SRAM, as shown in \Cref{fig:ipu-arch}. A typical ICCA chip example is 
Graphcore IPU~\cite{ipu2}. It has 1472 cores, 
each core has 624KB local SRAM and can access another core's SRAM at 5.5GB/s. 
This aggregates to an 896MB on-chip memory with 8TB/s all-to-all data exchange bandwidth.
The large on-chip space and high memory bandwidth present a promising way to break the memory wall for DL workloads (e.g., compared to an A100 GPU with 60MB total cache size and 2TB/s HBM bandwidth). With these advantages, 
inter-core interconnect has been employed by many AI chips today, such as Graphcore IPU~\cite{ipu2}, SambaNova SN40~\cite{samba-sn40}, Cerebras WSE~\cite{wse2}, Meta MTIA~\cite{meta-mtia}, and NVIDIA's H100 GPU~\cite{h100-hbm}. 

The inter-core interconnect connects all cores' local SRAM into a distributed memory space that can be managed by software (i.e., compiler),
leading to new parallel execution models for DL workloads (\mbox{$\S$\ref{sec:background:exec_model}}).
In conventional accelerators without inter-core connections, all cores execute independently with their local SRAM, and a separate global SRAM shared by all cores simultaneously handles all off-chip data loading.
On ICCA chips, the software can manually manage data sharing among cores without needing a global SRAM.
Also, the distributed nature of ICCA chip's on-chip SRAM allows its size to further scale, so it can store multiple tensor operators.
\hlcommon{Thus, when executing a current operator, the chip can simultaneously preload future operators' data from off-chip memory to SRAM.}
\hlF{However, this requires each core's local SRAM to enable double buffering between execution and preload, resulting in significant memory footprint overhead.}

\begin{figure}[t]
    \centering
    \vspace{1ex}
    \includegraphics[width=\linewidth]{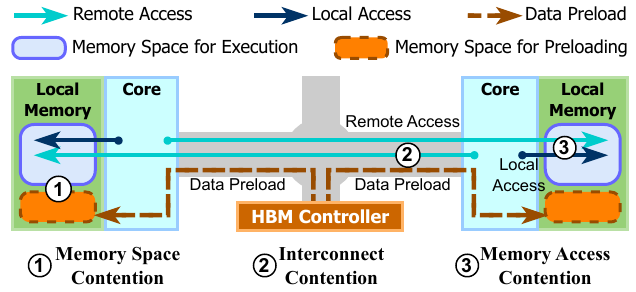} 
    \vspace{-3.5ex}
    \caption{Resource contentions on ICCA chip with HBM.}
    \label{fig:background:contention}
\end{figure}

The end-to-end performance of running a DL model on an ICCA chip is determined by three major factors: \textit{(1) compute (per-core execution), (2) communication (inter-core data exchange), and (3) I/O (data loading from off-chip memory)}.
To maximize the efficiency of the inter-core connected AI chip, it is challenging for software (i.e., DL compiler) to optimize all three performance factors, since they usually have conflicting resource demands, as shown in Figure~\ref{fig:background:contention}.




First, to overlap computing and off-chip loading, the DL compiler needs to decide how much \hlcommon{on-chip memory space} to allocate for per-core execution and for buffering preloaded data from off-chip memory.
\hlcommon{A larger space for execution (i.e., \textit{execution space}) allows larger per-core tile size, reduces inter-core communication traffic, and improves compute efficiency.
A larger space for preload (i.e., \textit{preload space}) improves off-chip memory bandwidth utilization.}
This leads to \textit{on-chip memory capacity contention} (\ding{172} in Figure~\ref{fig:background:contention}).
Second, the on-chip interconnect links all cores and HBM controllers, and its bandwidth is shared between inter-core data exchange and HBM-to-core data loading. This leads to \textit{interconnect bandwidth contention} (\ding{173}).
Third, the per-core SRAM must feed data to the local computation pipeline and serve data to other cores via the interconnect. The concurrent SRAM accesses will lead to \textit{memory access contention} (\ding{174}).


Given the performance trade-offs, we must jointly optimize all three performance factors. 
\hlB{However, to the best of our knowledge, few existing studies optimized the end-to-end performance by holistically considering all three performance factors
(i.e., per-core execution, inter-core data exchange, and off-chip data loading).}
Many DL compilers tune the tile size to optimize compute efficiency and off-chip memory access volume~\mbox{\cite{roller, soma}}, but do not consider the inter-core communication.
Some ICCA chip compilers like T10~\cite{t10} leverage new parallel execution paradigms to streamline the on-chip dataflow~\mbox{\cite{samba-sn40,tenstorrent,t10}}, which optimizes both per-core execution and inter-core communication. However, they did not consider the off-chip memory access.


In this paper, we present \pname{}, a DL compiler framework to maximize the efficiency of ICCA chips by jointly optimizing all the three performance factors. \pname{} formalizes these factors into a global trade-off space, based on the insight that these factors can be transferred into configurable compiler parameters ($\S$\ref{sec:motivation}), and the correlation between these parameters can reflect their performance trade-offs. 

Specifically, 
(1) the per-core execution performance is correlated to the SRAM capacity allocated to the \textit{execution space}.
(2) The off-chip data loading performance (i.e., the HBM bandwidth utilization) can be improved by increasing the \textit{number of preloaded operators}, which allows more overlap between computation and HBM access.
(3) An operator's inter-core data exchange overhead can be reduced by increasing the \textit{operator preload space}, which allows us to duplicate shared data in multiple cores in advance to avoid on-demand access to other cores, at the expense of higher SRAM footprint.


To search an optimized model execution plan, \pname{} schedules the preload and execution of each operator
with a two-level search algorithm 
to best overlap off-chip data access and on-chip execution.
For each operator, 
\pname{} first selects the optimal number of preloaded operators via an exhaustive search. 
The search space is small as the on-chip memory stores a limited number of preloaded operators.
Second, 
\pname{}'s cost-aware memory allocation algorithm determines the execution space size for the current operator and the preload space for each preloaded operator.
\pname{} uses an iterative greedy algorithm to minimize the execution time of the current operator and the inter-core data exchange overhead of preloaded operators.

As \pname{} preloads multiple operators, the earlier an operator is preloaded, the longer it occupies the on-chip SRAM, which limits the execution space of the current operator.
\hlcommon{Thus, \pname{} reorders the operator preloads to delay the preloads of operators that involve large tensors, reducing the lifespans of large operators' SRAM footprints.}
Also, as some operators require higher interconnect bandwidth to be preloaded to destination cores, 
\pname{} reorders the preload traffic to avoid ``rush hours'' on the interconnect, reducing the interconnect contention.
To yield an efficient search space of preload orders, \pname{} smartly limits the edit distance of preload orders based on the available SRAM capacity on the chip.


To evaluate \pname{}, we build an emulation framework using a real IPU-POD4 hardware~\cite{ipu_pod4} to emulate full-fledged ICCA chips with HBM, and a simulator framework with popular inter-core network topologies for sensitivity analysis and design space exploration of ICCA chips. 
We evaluate \pname{} with state-of-the-art LLMs and stable diffusion models. 
We not only show \pname{} achieves 94\% of the ideal roofline performance, but also present \pname{}'s capability of exploring design tradeoffs in ICCA chips. 
We list our contributions as follows:


\begin{itemize} [leftmargin=*]

    \item For the inter-core connected AI chip with off-chip HBM, we are the first to identify the performance challenges for best utilizing its hardware properties.
    
    \item We develop a DL compiler framework \pname{} that structures the performance factors into configurable parameters in the compiler, 
    such that we can optimize hardware performance by exploring the space using compiler techniques.
    
    \item 
    We develop a new inductive operator scheduling policy in \pname{} for optimizing the overlapping of HBM data loading and on-chip execution, as well as design a new cost-aware algorithm for on-chip memory allocation.
    \item To generalize our design in the DL compiler, we build a generic interface that can map the optimized end-to-end execution plan to popular ICCA chip architectures. 
    \item To evaluate our design, we construct an emulation framework with real IPU-POD4~\cite{ipu_pod4} hardware, and demonstrate the efficiency of \pname{} for various DL models. 
    \item We build the first hardware simulator for ICCA chips, which supports popular network topologies for inter-core communications and various bandwidth behaviors.
    \item With \pname{} and the ICCA chip simulator, we enable design space exploration of ICCA chips and present our insights in $\S$\ref{sec:eval:sens}. We will open source our codebase to the community.  
\end{itemize}

%% file: background.tex
\section{Background and Motivation}
\label{sec:background}

We now introduce the features of the inter-core connected AI chip and discuss the motivation of \pname{}.

\subsection{Architecture of the ICCA Chip}
\label{sec:background:ipu-arch}

To facilitate the introduction of the ICCA chip as shown in \Cref{fig:ipu-arch}, we use Graphcore IPU MK2~\cite{ipu2} as an example.
An IPU chip has 1472 cores that execute independently in parallel. 
Each core has 624KB local scratchpad memory, adding up to 896MB of total on-chip memory.
All cores are interconnected with high-bandwidth low-latency links. Each core can access any other core's local memory at 5.5GB/s, delivering an aggregated inter-core all-to-all bandwidth of $1472\times5.5\text{GB/s}\approx 8\text{TB/s}$~\cite{ipu_citadel}.
The large on-chip memory improves on-chip data reuse by storing more operators or even an entire model. 
The all-to-all interconnect allows each core to independently access on-chip data at high bandwidth.
If multiple cores receive/send different data from/to the same core, the interconnect sequentially serves each data transfer at full bandwidth.
 In addition to IPU, other ICCA chips, such as SambaNova SN40L~\mbox{\cite{samba-sn40}} and Tenstorrent~\mbox{\cite{tenstorrent}}, feature a mesh-based on-chip interconnect.
In general, the ICCA chip architecture enables scalable performance and alleviates the memory bandwidth bottleneck for serving memory-intensive DL workloads like LLMs.

\noindent
\textbf{Scale the ICCA chip with HBM.}
To serve large models whose sizes exceed the on-chip capacity, we can scale the memory capacity 
with off-chip memory modules like high bandwidth memory (HBM)~\cite{hbm3e}.
Many ICCA chips already integrate off-chip memory~\cite{samba-sn40,meta-mtia}. 
As shown in \Cref{fig:ipu-arch}, they attach HBM controllers to the on-chip interconnect, so each controller can directly send data to each core similar to how the cores send data to each other.
To access HBM, cores communicate with HBM controllers via the interconnect.
The HBM controller coalesces the memory requests from cores, loads data from HBM, and sends data to cores.



\subsection{Execution Model of ICCA Chip with HBM}
\label{sec:background:exec_model}


Before executing a DL model, all required data (e.g., model weight) is loaded into HBM.
The ICCA chip will sequentially execute each operator in the model by first preloading its required data from HBM to on-chip memory and then performing the on-chip computation.
To maximize computing throughput,
the compiler manages the on-chip SRAM as a double buffer to overlap the on-chip execution and off-chip HBM access.
The compiler partitions the SRAM in each core into an \textit{execution space} to store the currently executing operator and a \textit{preload space} to store the operators preloaded from HBM.
While an operator is executing, the ICCA chip can preload other operators from the HBM into the on-chip memory.
On preload, HBM controllers use the interconnect to deliver preloaded data to cores.
Each core needs to reserve enough local memory space for this.
When the preload space is full, the preload will stop and the HBM bandwidth will be underutilized.

\begin{figure}[t]
    \centering
    \vspace{0.3ex}
    \includegraphics[width=1\linewidth]{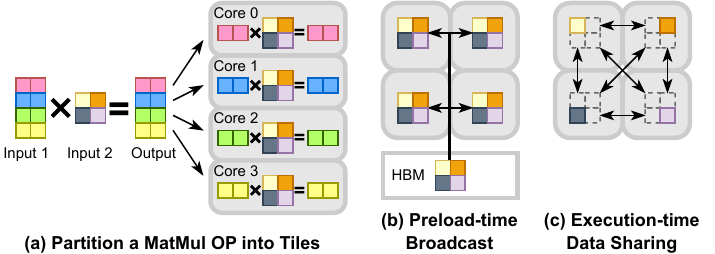}
    \vspace{-4ex}
    \caption{Operator partitioning and inter-core data sharing.}
    \label{fig:background:sharing-types}
    \vspace{-0.5ex}
\end{figure}

\noindent
\textbf{On-chip execution.}
Several parallel execution models can execute tensor operators on ICCA chips~\mbox{\cite{roller,t10,megatron}}, all of them  require significant computation, on-chip memory, and communication resources.
In these execution models, a compiler will partition the computation of a tensor operator into small tiles~\cite{roller,t10} and map each tile to a core.
To execute a tile, each core must fetch the required data from HBM or another core to its local memory, via the interconnect.
For example, a MatMul operator is partitioned into four tiles in \Cref{fig:background:sharing-types} (a),
and all cores require ``Input 2'' for per-core execution.
In some execution models~\mbox{\cite{megatron}}, this shared tensor will be directly broadcast to each core by the HBM controller via the interconnect, as shown in \mbox{\Cref{fig:background:sharing-types}} (b). This needs more local memory space but fewer inter-core accesses.
Some other execution models~\mbox{\cite{t10, roller}} allows a core to access shared tensors from other cores during execution, as shown in \mbox{\Cref{fig:background:sharing-types}} (c).
This needs less per-core local space but more inter-core accesses.
After the per-core execution, an operator may need to reduce the partial results across multiple cores into the final result, where these cores will exchange the partial results via the interconnect.

\subsection{Challenges of Using ICCA Chip with HBM}
\label{sec:background:challenges}


To maximize the ICCA chip performance, we must (1) allow faster per-core execution, (2) utilize more HBM bandwidth to preload required data on time, and (3) reduce the inter-core data sharing overhead.
However, it is difficult to maximize all three performance metrics simultaneously, as they have conflicting resource demands.

\noindent
\textbf{On-chip memory space contention.}
We cannot \textit{maximize per-core execution performance} and \textit{HBM bandwidth utilization} at the same time, due to on-chip memory space contention.
As shown by \ding{172} in \Cref{fig:background:contention},
each core reserves an execution space for the currently executing operator and a preload space for the preloaded operators.
To speed up per-core execution, a larger execution space is required (see \S\ref{sec:background:fast-exec} and \Cref{fig:op-exec-time}).
To prevent HBM underutilization, a larger preload space is required (see \S\ref{sec:background:hbm-util} and \Cref{fig:sram-hbm}).
With limited on-chip memory, we cannot expand both spaces.

\noindent
\textbf{Interconnect bandwidth contention.}
We cannot \textit{maximize HBM bandwidth utilization} and \textit{minimize inter-core data sharing overhead} at the same time, due to the interconnect bandwidth contention.
As shown by \ding{173} in \Cref{fig:background:contention}, the on-chip interconnect carries both core-to-core traffic for inter-core data sharing and HBM controller-to-core traffic for preloading.
When both traffic flows are heavy, the interconnect will be congested (see \S\ref{sec:background:interconnect-util} and \Cref{fig:shift-ratio-all}).

\noindent
\textbf{Memory access contention.}
We cannot \textit{maximize per-core execution performance} and \textit{minimize inter-core data sharing overhead} at the same time, due to the memory access contention.
As shown by \ding{174} in \Cref{fig:background:contention}, each core's local memory is simultaneously accessed by the core itself for computing a tile, and by other cores for inter-core data sharing.
For example on IPU, each core reads its local memory at full speed (128 bits/cycle~\cite{ipu_isa}) when executing DL operators like MatMul, any other accesses will pause the execution.
Upon contention, tile execution on this core reads data from local memory at slower speed, or even pauses entirely. The remote cores may also suffer from degraded SRAM bandwidth.

%% file: motivation.tex
\vspace{-0.1em}
\section{Performance Tradeoffs in \pname{}}
\label{sec:motivation}
\vspace{-0.1em}

\begin{figure}[t]
    \centering
    \includegraphics[width=\linewidth]{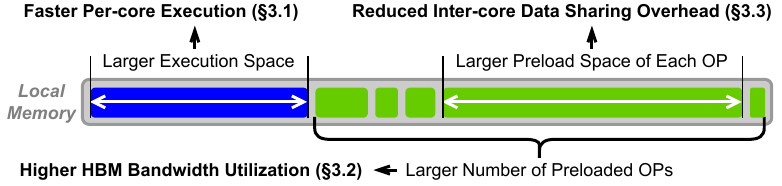} 
    \vspace{-3.5ex}
    \caption{Mapping performance factors to compiler decisions on per-core SRAM allocation among preload and execution.}
    \label{fig:design:tradeoff}
\end{figure}

As discussed in $\S$\ref{sec:background:challenges}, to efficiently use ICCA chips with HBM, we must trade-off multiple performance factors.
We summarize how each performance factor is mapped to a compiler decision in \Cref{fig:design:tradeoff}.
First, increasing per-core \textit{execution space} enables faster per-core execution with a larger tile size (\S\ref{sec:background:fast-exec}).
Second, increasing \textit{number of preloaded operators} can better overlap on-chip execution and off-chip HBM load, improving HBM bandwidth utilization (\S\ref{sec:background:hbm-util}).
{Third, increasing \textit{preload space} for each preloaded operator reduces the inter-core data exchange overhead and the memory access contention, since {the shared data can be duplicated on cores in advance to reduce the overhead of on-demand accesses to other cores (\mbox{\S\ref{sec:background:interconnect-util}})}.}
We validate the insights with experiments on our ICCA chip emulator (see implementation in $\S$\ref{sec:implt}) as follows.

\begin{figure}[t]
    \centering
    \includegraphics[width=\linewidth]{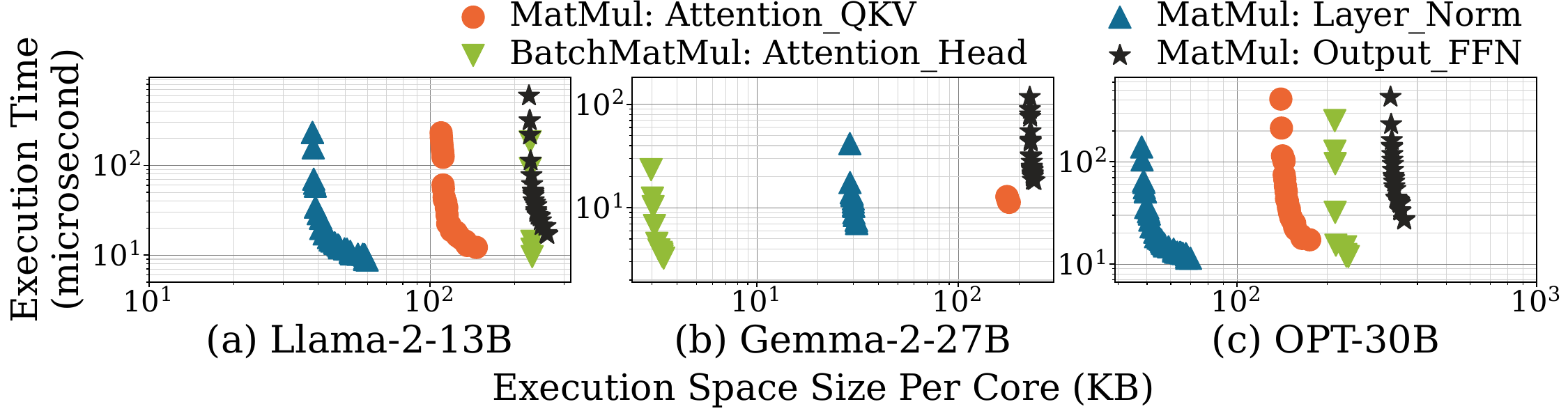} 
    \vspace{-3.5ex}
    \caption{The execution times of representative operators given different per-core execution spaces. Each data point is a plan. In each model, plans of the same operator use the same legend (e.g., \textit{MatMul:Attn\_QKV} is the MatMul operator that calculates the Q,K,V matrices in attention~\cite{attentionisallyouneed}).}
    \label{fig:op-exec-time}
\end{figure}

\subsection{Larger Execution Space Enables Faster Per-core Execution}
\label{sec:background:fast-exec}
\vspace{-0.15em}


There are many possible ways to partition an operator~\cite{roller,ansor,alpa}, resulting in partition plans with different per-core tile sizes, execution times, and inter-core data exchange traffic.
Generally, a larger execution space per core enables a larger tile size and improves the per-core execution performance,
as a larger tile implies higher per-core data reuse and larger compute granularity, with fewer inter-core data accesses.
We show the correlation between execution time and execution space in \Cref{fig:op-exec-time}. 
We choose representative operators from popular LLMs of various sizes: Llama-2-13B~\cite{llama2}, Gemma-2-27B~\cite{gemma}, and OPT-30B~\cite{opt}.
For each operator, we plot the execution times of partitioning plans generated by a state-of-the-art compiler~\cite{t10} for ICCA chips given different SRAM size constraints.
The results show that faster execution plans require more per-core execution space. We observe similar results in DL compilers using other parallel execution models~\mbox{\cite{roller, ansor}}.

Existing compilers focus on achieving higher on-chip execution performance using a given execution space size.
However, they cannot find a proper execution space size by arbitrating the memory space contention in \mbox{\S\ref{sec:background:challenges}}.
Moreover, operators from the same model have diverse memory vs. time correlations.
Thus, we also need to adjust the execution space size based on each operator's performance characteristic, rather than allocating a fixed-sized execution space throughout the model execution.



\subsection{Preloading More Operators Improves HBM Bandwidth Utilization}
\label{sec:background:hbm-util}
Operators in a DL model have different \textit{compute intensities} (i.e., number of floating-point operations, or FLOPs, performed per byte).
While some are compute-intensive due to more on-chip data reuse (e.g., operators that use model parameters, which are reused by all input requests in a batch), others are memory-intensive (e.g., the KV cache~\cite{efficient_scale_trans_infer}, which has no data reuse among requests in a batch).

The diverse HBM access and execution time across operators cause sub-optimal computation and HBM bandwidth utilization.
If the currently executing operator has a short execution time while the next operator has a long HBM time, the current operator finishes before the next operator completes preloading, and the computation stalls.
Similarly, if the next operator finishes preloading before the current operator completes, the HBM bandwidth is underutilized.




\begin{figure}[t]
    \centering
    \includegraphics[width=\linewidth]{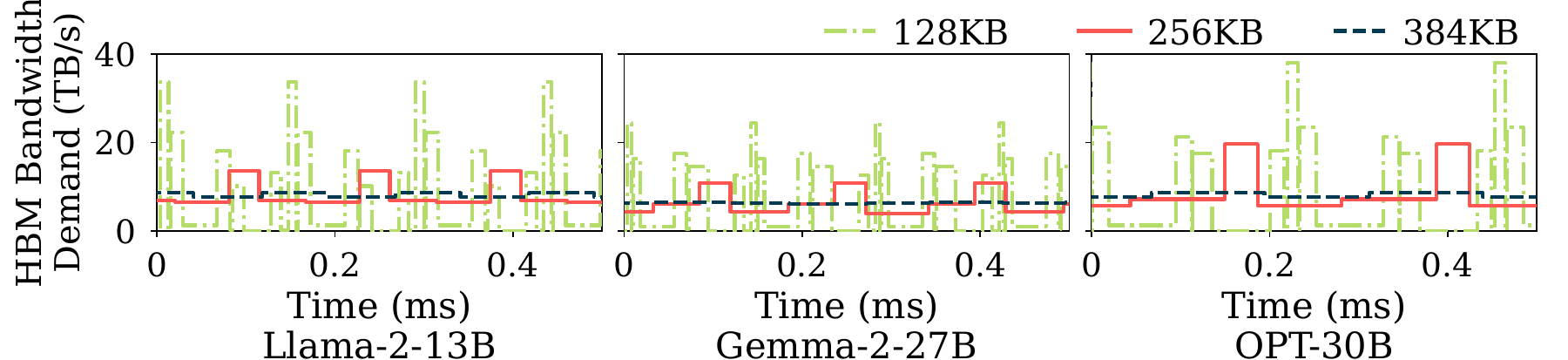} 
    \vspace{-3ex}
    \caption{HBM bandwidth demands of models across time, given different preload spaces. The legend shows per-core preload space size in KB (same for all cores).}
    \label{fig:sram-hbm}
\end{figure}

To improve HBM bandwidth utilization, we can preload more operators. 
This also improves compute utilization, as more data will be ready on-chip, so future execution is less likely to stall.
However, preloading more operators requires a larger preload space.
\Cref{fig:sram-hbm} shows how the HBM bandwidth demand varies over time for LLM inference with different per-core preload space sizes.
The bandwidth demand is quantified as the minimum HBM bandwidth to prevent on-chip execution from stalling.
With small preload space, the bandwidth demand fluctuates drastically due to insufficient preload opportunities.
With larger preload space, more operators can be preloaded. This smooths out the bandwidth demand, reduces the compute/memory idleness, and enhances the overall performance. 

\begin{figure}[t]
    \centering
    \includegraphics[width=1\linewidth]{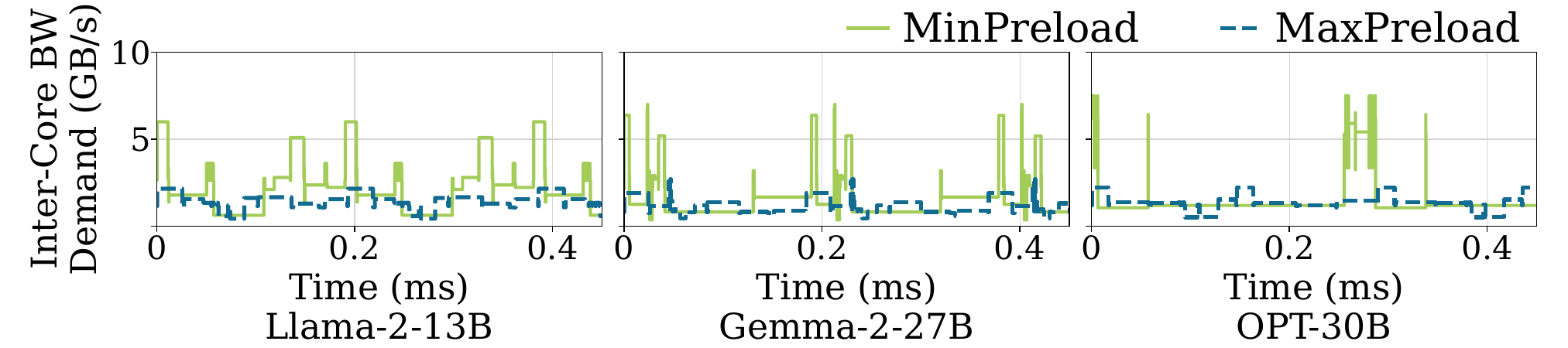}
    \vspace{-4ex}
    \caption{The inter-core bandwidth demand of each core 
    across time, with different preload settings.
    The demand does not count HBM controller-to-core traffic.}
    \label{fig:shift-ratio}
\end{figure}

\begin{figure}[t]
    \centering
    \includegraphics[width=\linewidth]{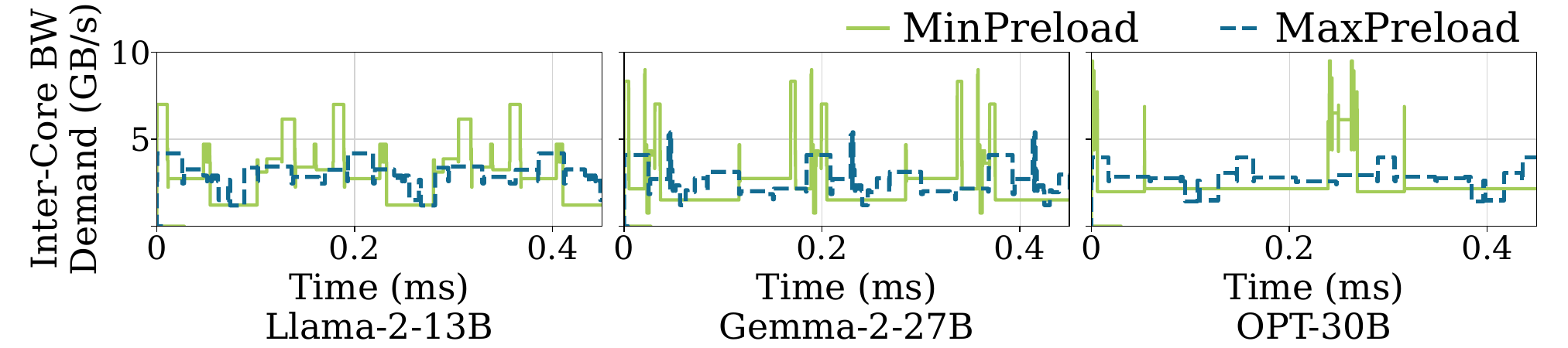}
    \vspace{-4ex}
    \caption{The total per-core interconnect bandwidth demand.}
    \label{fig:shift-ratio-all}
\end{figure}


\subsection{Larger Per-Operator Preload Space Reduces Inter-core Data Access Volume}
\label{sec:background:interconnect-util}

As discussed in \S\ref{sec:background:exec_model}, data shared between cores can be either broadcasted by HBM controllers during preload or accessed from peer cores during execution.
A larger preload space allows for more broadcasts at preload time and fewer on-demand accesses at execution time.
{Also, with fewer inter-core accesses, less memory access contention will occur on each core.}

\Cref{fig:shift-ratio} shows that expanding preload space reduces the inter-core bandwidth demand.
For each operator, we pick the fastest execution plan that fits in a given execution space size\footnote{For each run, we use the optimal execution space size that gives the smallest total inference latency. See the description of the \baseline{} setup in \S\ref{sec:eval:setup}.}.
MinPreload lets each core access all shared data from other cores at execution time, which requires the minimum preload space. MaxPreload lets HBM controllers broadcast as much shared data as possible at preload time, which requires the largest preload space.
We profile the inter-core bandwidth demand ($\frac{\text{inter-core transfer volume}}{\text{per-core execution time}}$) of each core. MaxPreload significantly reduces the inter-core traffic.

Although more broadcasts on preload increase the HBM controller-to-core traffic, the preload traffic can be opportunistically interleaved with ongoing inter-core traffic to reduce contention.
\Cref{fig:shift-ratio-all} shows how each core's total interconnect bandwidth demand (defined as $\vspace{-0.2em}\frac{\text{inter-core transfer volume}}{\text{per-core execution time}}+\frac{\text{HBM-to-core transfer volume}}{\text{HBM load time}}\vspace{0.1em}$) varies over time.
Purely relying on inter-core transfer fluctuates the traffic pressure drastically, causing interconnect underutilization or congestion.
More broadcasts at preload time reduce fluctuation by spreading the traffic across preload and execution times.

%% file: design.tex
\section{Design and Implementation}
\label{sec:design}

\begin{figure}[t]
    \centering
    \includegraphics[width=1\linewidth]{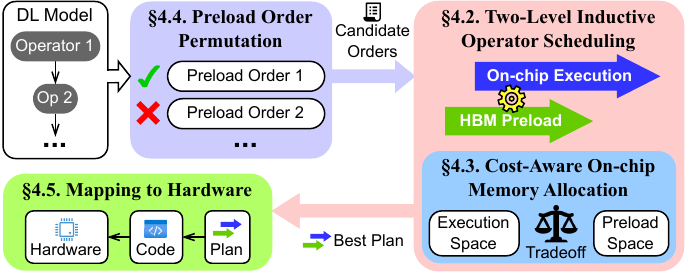} 
    \vspace{-3.5ex}
    \caption{Overview of our \pname{} framework.}
    \label{fig:design:overview}
\end{figure}

We design \pname{}, a compiler framework for exploring the efficiency of ICCA chips. \pname{} automatically trades-off performance factors by configuring the number of preloaded operators, the per-core execution space size, the per-operator preload space size, and the preload order of operators.
\hlcommon{We show the design overview of \pname{} in \mbox{\Cref{fig:design:overview}}. We use \mbox{\Cref{tab:relation}} to show which design component of \pname{} handles each performance tradeoff in \mbox{\S\ref{sec:motivation}}.}

\vspace{-0.2em}
\subsection{Design Overview}
\label{sec:design:overview}


For a DL model,
\pname{} schedules the preload and execution of operators by exploring a two-level search space.
\hlcommon{First, for each operator, \pname{} explores all possible numbers of future operators to preload before or during this operator's execution (\mbox{\S\ref{sec:design:algorithm}}).}
Second, for each number of preload operators, \pname{} optimizes on-chip memory allocation by trading off between execution and preload spaces
(\S\ref{sec:design:allocation}).

\renewcommand{\tabcolsep}{2pt}
\begin{table}[t]
    \centering
    \caption{
        \hlcommon{A summary of performance tradeoffs (\mbox{\S\ref{sec:motivation}}) investigated in our design (\mbox{\S\ref{sec:design}}).}
    }
    \vspace{-.5ex}
    \footnotesize
    \begin{tabular}{|c|c|c|}
    \hline
        \textbf{Compiler Decision} & \textbf{Relevant Performance Factors} & \textbf{Relevant Design} \\\hline
        \specialcell{Number of operators\\to preload ahead}
        & \specialcell{(1) Improve HBM bandwidth utilization}
        & \specialcell{Two-level inductive\\scheduling (\S\ref{sec:design:algorithm})} \\\hline
        \specialcell{Execution space size}
        & \specialcell{(1) Accelerate per-core execution \\(2)  Reduce inter-core data accesses\\(i.e., reduce interconnect and\\memory access contentions)} 
        & \specialcell{Cost-aware memory\\allocation (\S\ref{sec:design:allocation})} \\\hline
        \specialcell{Preload space size of\\each operator}
        & \specialcell{(1) Reduce inter-core data accesses\\(i.e., reduce interconnect and\\memory access contentions)}
        & \specialcell{Cost-aware memory\\allocation (\S\ref{sec:design:allocation})} \\\hline
        \specialcell{Preload order}
        & \specialcell{(1) Reduce interconnect contention \\(2) Reduce the lifespans of large\\operators' preload spaces} 
        & \specialcell{Preload order\\permutation (\S\ref{sec:design:reorder})} \\\hline
    \end{tabular}
    \label{tab:relation}
    \vspace{0.5ex}
\end{table}

To reduce the inter-core data exchange overhead and enable larger execution space, \pname{} allows operators to be preloaded in a different order. 
\pname{} finds the optimal preload order by searching through all promising orders.
For each order, \pname{} applies operator scheduling policies and conducts a performance estimation.
To reduce the search overhead, \pname{} prunes orders that will overflow the on-chip memory (\S\ref{sec:design:reorder}).
Finally, \pname{} generates an optimized end-to-end plan for the entire model.
The plan specifies the preload and execution plan of each operator.
A code generator then translates this plan into an executable program for the hardware
(\S\ref{sec:design:together}).

\vspace{-0.2em}
\subsection{Two-Level Inductive Operator Scheduling} 
\label{sec:design:algorithm}





\newcommand{\T}[3]{T^{#1}_{\textit{{#2}-{#3}}}}
\newcommand{\Tspre}[1]{\T{#1}{s}{pre}}
\newcommand{\Tepre}[1]{\T{#1}{e}{pre}}
\newcommand{\Tsexe}[1]{\T{#1}{s}{exe}}
\newcommand{\Teexe}[1]{\T{#1}{e}{exe}}
\newcommand{\Tstart}{T_{\textit{start}}}
\newcommand{\Tend}{T_{\textit{end}}}

The scheduling algorithm minimizes the end-to-end execution time of a DL model by deciding the number of future operators to preload before or during each operator's on-chip execution (i.e., \textit{preload number}).
Each preload number represents a trade-off point between on-chip execution speed and HBM bandwidth utilization (\Cref{fig:design:tradeoff}).

The optimization space is exponential to the number of operators.
Suppose there are $N$ operators in a DL model, and the on-chip memory can fit at most $K$ operators. Each operator's execution can overlap with $1$ to $K$ operators' preload. Thus, there are $O(K^N)$ combinations of preload numbers for all operators. For example, for IPU-POD4 (3.5GB on-chip memory) and OPT-30B, each identical layer has 84 operators and $K\ge28$, forming up to $28^{84}$ combinations.

We develop an $O(KN)$-time algorithm for this problem.
The insight is, as operators in a DL model typically execute in a sequential order due to data dependency, instead of exploring all combinations of preload numbers, we can exploit the execution order and inductively derive the optimal preload number for each operator.

We can either start from the first operator and find the optimal preload number for each succeeding operator, or start from the last operator and schedule each preceding operator.
For each operator, we explore all possible preload numbers based on the already scheduled operators, and we pick the preload number that minimizes the ``start-to-current'' or ``current-to-end'' execution time.
As both induction directions are equivalent, we focus on the second one.

\begin{figure}[t]
    \centering
    \vspace{-0.6ex}
    \includegraphics[width=\linewidth]{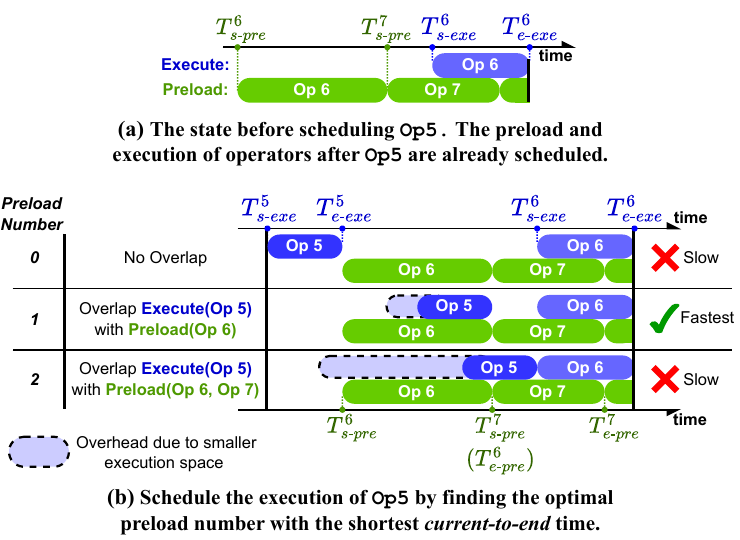} 
    \vspace{-3.5ex}
    \caption{Select the preload number that minimizes the ``current-to-end" time.}
    \label{fig:design:overlap}
    \vspace{-.5ex}
\end{figure}

The base case of induction is trivial, as the last operator has no succeeding operators to preload (i.e., preload number is always 0). For the inductive step, we show an example in \Cref{fig:design:overlap}.
In \Cref{fig:design:overlap} (a), \pname{} has finished scheduling all operators after \textstt{Op5}.
Then, \pname{} schedules the execution of \textstt{Op5} in \Cref{fig:design:overlap} (b).
\pname{} enumerates all possible preload numbers for \textstt{Op5}.
For each preload number, \pname{} invokes the cost-aware on-chip memory allocation algorithm ($\S$\ref{sec:design:allocation}) to determine the execution/preload space sizes for the involved operators and the estimated execution time of \textstt{Op5}.

For example, preload number 0 means we do not overlap \textstt{Op5}'s execution with any preload. The execution time of \textstt{Op5} is minimized, but the overall execution time from \textstt{Op5} to the end of the model is sub-optimal.
For preload numbers 1 and 2, \pname{} overlaps the execution of \textstt{Op5} with the preload of \textstt{Op6}, or the preloads of both \textstt{Op6} and \textstt{Op7}. Though \textstt{Op5}'s execution time is longer, the overall execution times are better than preload number 0. As preload number 1 yields the lowest current-to-end time, \pname{} selects it for \textstt{Op5}.

After scheduling \textstt{Op5}'s execution, \pname{} schedules its preload to occur just before its execution or before \textstt{Op6}'s preload, whichever is earlier, to preserve data dependency.
Scheduling the preceding operator (\textstt{Op4}) will depend on \textstt{Op5}'s preload time, which is estimated as the maximum of (1) the HBM access time from a roofline model~\cite{roofline} and (2) the interconnect transfer time from the cost model in \S\ref{sec:design:allocation}.

Our algorithm has $O(KN)$ complexity as we iterate through $N$ operators with up to $K$ preload numbers per operator.
The algorithm provably finds the end-to-end plan with the shortest total time, assuming it can obtain the optimal execution time for each preload number.
Lemma \ref{lemma:basecase} and \Cref{thm:induction} formalize the algorithm.

\renewcommand{\qedsymbol}{$\blacksquare$}

\vspace{-1ex}
\begin{lemma}[Base case]\label{lemma:basecase}
    Given a model with $N$ operators, for each operator $i$, let $\Tspre{i}$ and $\Tepre{i}$ be the start and end time of operator $i$'s preload.
    Let $\Tsexe{i}$ and $\Teexe{i}$ be the start and end time of operator $i$'s execution.
    Let $\Tstart = \Tspre{1}$ and $\Tend = \Teexe{N}$ be the start and end time of the model execution.
    Then, for operator $N$, preload number 0 minimizes $\Tend - \Tsexe{N}$.
\end{lemma}
\vspace{-1ex}

\begin{proof}
Since operator $N$ is the last operator, the only possible preload number is 0.
\end{proof}


\newtheoremstyle{noindenttheorem}
{}                
{}                
{\itshape}        
{}                
{\bfseries}       
{.}               
{ }               
{}                
\theoremstyle{noindenttheorem}
\vspace{-1ex}
\begin{theorem}[Inductive step]\label{thm:induction}
    Let $1$$\leq$$i$$<$$N$. Suppose we have minimized $\Tend - \Tsexe{i+1}$.
    Then, there exists a preload number $p$ whose $\Tsexe{i}$ minimizes $T_{end}-\Tsexe{i}$, or maximizes $\Tsexe{i}$.
    Specifically, we have $\Teexe{i} = \min(\Tsexe{i+1}, \Tspre{i+p+1})$, and $\Tsexe{i}=\Teexe{i}-L^i_{exe}$ where $L^i_{exe}$ is the execution time of operator $i$ derived by the cost-aware memory allocation algorithm in $\S$\ref{sec:design:allocation}. 
\end{theorem}
\vspace{-1.5ex}

\begin{proof}
    First, to prove $\Teexe{i} = \min(\Tsexe{i+1}, \Tspre{i+p+1})$ for any preload number $p$,
    we have (1) \textstt{Op}$i$ must finish execution before \textstt{Op}$(i+1)$ starts execution, e.g., $\Teexe{i} \leq \Tsexe{i+1}$; and (2) \textstt{Op}$i$'s execution can be overlapped with the preload of the next $p$ operators, which implies \textstt{Op}$i$'s execution must finish before the preload of \textstt{Op}$(i+p+1)$, e.g., $\Teexe{i} \leq \Tspre{i+p+1}$.
    Next, we prove the existence of $\Tsexe{i}$ that minimizes $T_{end}-\Tsexe{i}$. Suppose by contradiction that $T_{end}-\Tsexe{i+1}$ is minimized but there is no $\Tsexe{i}$ that minimizes $T_{end}-\Tsexe{i}$.
    Since our inductive step explored all $\Tsexe{i}$ values by enumerating all possible preload numbers,
    the only possible case is that we must explore more preload numbers to find the global $\max(\Tsexe{i})$, or the global $\max(\Teexe{i})$ is greater than $\Tsexe{i+1}$, which means $\Tsexe{i+1}$ can be larger.
    This is a contradiction since $T_{end}-\Tsexe{i+1}$ is already minimized, e.g., $\Tsexe{i+1}$ is already maximized.
\end{proof}

\subsection{Cost-Aware On-chip Memory Allocation}
\label{sec:design:allocation}

In \S\ref{sec:design:algorithm}, when scheduling an operator, \pname{} optimizes the performance for each preload number.
Given the currently executing operator and a set of operators to be preloaded, \pname{} defines a \textit{two-level tradeoff space between the execution/communication time and the memory consumption}.

First, there are two types of \textit{\textbf{intra-operator tradeoffs}}:
(1) For the currently executing operator, \pname{} trades memory space for execution time ($\S$\ref{sec:background:fast-exec}).
(2) For each preloaded operator, \pname{} trades memory space for this operator's inter-core data exchange overhead ($\S$\ref{sec:background:interconnect-util}).
Second, there is an \textit{\textbf{inter-operator tradeoff}:}
as operators have different memory-time tradeoffs, we allocate more memory to operators that benefit more from a larger execution/preload space.

\pname{} explores the two-level space in two stages.
First, for each operator, \pname{} finds all Pareto-optimal tradeoff plans between time and memory.
Second, \pname{} jointly determines the execution/preload space sizes of all operators based on the Pareto-optimal plans and the total on-chip memory capacity.

\begin{figure}[t]
    \centering
    \vspace{0.5em}
    \includegraphics[width=\linewidth]{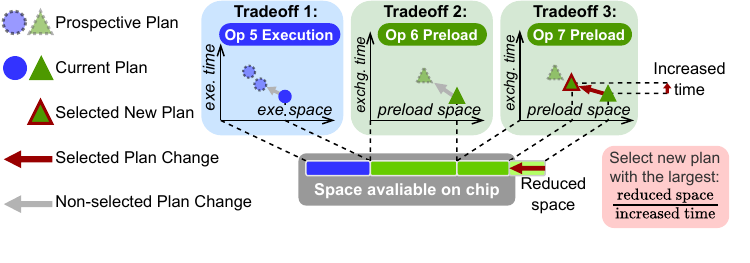} 
    \vspace{-7.5ex}
    \caption{Tradeoff between time overhead \& memory usage.}
    \label{fig:design:allocation}
    \vspace{-1ex}
\end{figure}

\noindent
\textbf{\textul{Intra-operator tradeoff for on-chip execution}} (Tradeoff 1 in \Cref{fig:design:allocation})\textbf{.}
For the currently executing operator, there are many \textit{partition plans} to partition its computation into tiles, each runs on one core (\Cref{fig:background:sharing-types}).
\hlB{\pname{} integrates existing compiler techniques to enumerate all partition plans of an operator given its operator type and tensor shapes \mbox{\cite{roller, ansor, tensorir, samba-sn40, t10, tenstorrent-TT-METALIUM}}.
These techniques represent each plan as a list of integers (see examples in \mbox{\S\ref{sec:implt}}) and check if a plan is compatible with the target hardware (e.g., not using more cores than available, not overflowing the SRAM).}
For each plan,
\pname{} estimates its execution time using a cost model and its execution space using the tile size.
\pname{} examines all plans to find the ones on the Pareto-optimal curve, where each plan either runs faster than any other plans that use the same or less memory, or uses less memory than any others with the same or less execution time.

\begin{figure}[t]
    \centering
    \includegraphics[width=\linewidth]{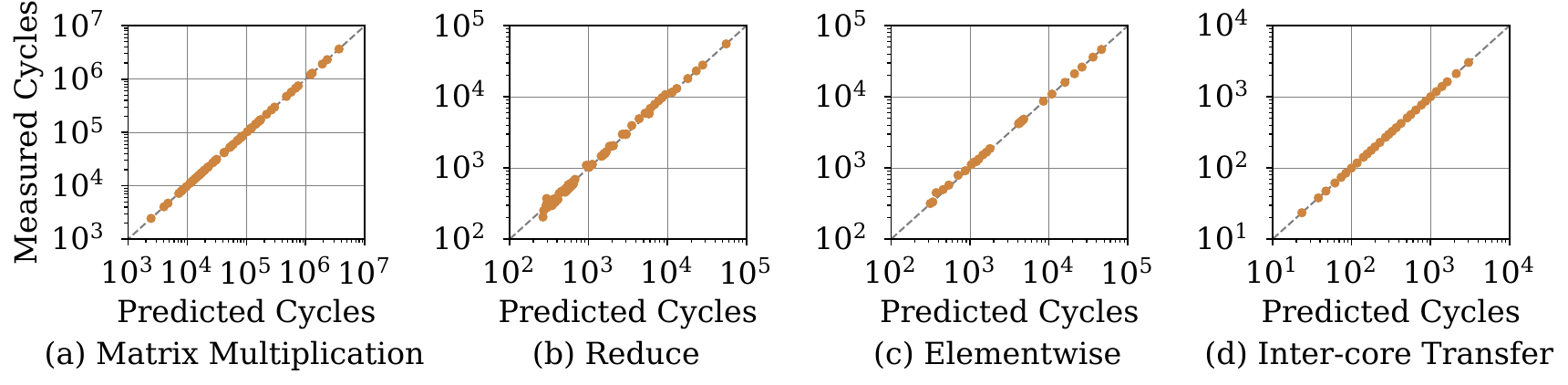} 
    \vspace{-4ex}
    \caption{Cost model accuracy of different operators and inter-core transfer, for different tile shapes. Each point is the measured vs. predicted per-core execution or transfer time.}
    \label{fig:design:cost}
\end{figure}

\noindent
\textul{\textit{Cost model for execution time.}}
As DL workloads have predictable execution patterns~\mbox{\cite{tensorir, Galvatron, roller, t10, vidur, regate}},
\pname{} uses an accurate cost model to quickly estimate the performance of per-core execution and inter-core transfer.
For each operator type (e.g., MatMul), we randomly generate tiles with varied shapes,
and run each tile using one core on the target device. 
\hlB{Then, we fit a linear tree model \mbox{\cite{linear-tree}} using the tile shapes as inputs and the profiled execution times as outputs.}
For inter-core transfer, we fit a model for each network link using transfer volumes as inputs and transfer times as outputs.
For each partition plan, \pname{} determines the tile-to-core mapping and orchestrates the inter-core transfer (e.g., the source/destination cores and intermediate hops of each transfer, 
see \mbox{$\S$\ref{sec:implt}}).
\pname{} uses the per-link cost model and the communication pattern to estimate the total transfer time.
\Cref{fig:design:cost} shows that \pname{} can accurately predict the execution and transfer times of an IPU chip.
{\pname{} can use different cost models~\mbox{\cite{Galvatron,roller,t10,vidur}} for different hardware platforms.}

\noindent
\textbf{\textul{Intra-operator tradeoff for preloading}} (Tradeoff 2 and 3 in \Cref{fig:design:allocation})\textbf{.}
For each preloaded operator, its partition plan is already decided in a previous step of the inductive operator scheduling ($\S$\ref{sec:design:algorithm}).
This \textit{execute-state plan} is chosen for execution speed, which may use more memory space.
As the operator is not currently executing, \pname{} assigns a memory-efficient \textit{preload-state plan}.
To start execution, a \textit{data distribution phase} transforms the operator from preload- to execute-state by distributing the required data via the interconnect (e.g., \Cref{fig:background:sharing-types} (c)).
It saves this operator's preload space at the cost of extra inter-core data exchange overhead, compared to broadcasting the required data at preload time following the execute-state plan (e.g., \Cref{fig:background:sharing-types} (b)).

Each execute-state plan may have many preload-state plans, by configuring how much data is broadcasted on preload.
On preload, if 4 cores share a data piece, we can evenly split it into 1, 2, or 4 chunks, and broadcast each chunk to 4, 2, or 1 cores.
Each core receives $1$, $\frac{1}{2}$, or $\frac{1}{4}$ of the data on preload (this decides preload space size), and fetches the rest $0$, $\frac{1}{2}$, or $\frac{3}{4}$ on data distribution.
\pname{} finds the Pareto-optimal preload-state plans of each preloaded operator, by estimating their preload space sizes and data distribution times.



\noindent
\textbf{\textul{Inter-operator tradeoff.}}
With limited on-chip memory, \pname{} jointly trades off memory allocation among the executing and preloaded operators.
\hlcommon{It minimizes the total time, which is determined by (1) execution times, (2) data-distribution times, (3) interconnect contention overhead due to overlapped preload and execution, and (4) memory access contention overhead between local SRAM accesses and inter-core accesses\footnote{\hlcommon{For some ICCA chips where local SRAM accesses are blocked by inter-core accesses (e.g., IPU), we estimate access contention overhead using the inter-core access time.}}.}
To estimate the contention overhead on each interconnect link, \pname{} divides total traffic by link bandwidth.

As enumerating all possible plan combinations is impractical (e.g., $O(P^K)$ combinations for $K$ operators each with $P$ plans), \pname{} uses a heuristic based on each operator's memory-cost efficiency.
\pname{} starts with each operator's fastest plan as the currently selected plan.
This combination of plans requires the most execution/preload space, so the total space requirement may exceed the memory capacity.
\pname{} then iteratively searches for the best combination of plans whose total memory requirement can fit into the on-chip memory,
at the cost of slightly increasing the total execution time.

For each search step, \pname{} examines the next plan with a smaller memory footprint along the Pareto-optimal curve for each operator.
\pname{} selects the most ``cost-effective'' operator whose next plan has the largest ratio $\Delta=\frac{\text{reduced space size}}{\text{increased time}}$ compared to the currently selected plan.
For example, in \Cref{fig:design:allocation}, \textstt{Op5} is the executing operator. \textstt{Op6} and \textstt{Op7} are preloaded operators.
\pname{} updates the current plan for \textstt{Op7} and proceed to the next search step.
\pname{} stops when the total memory requirement does not exceed the available capacity. 




In the worst case, \pname{} needs to examine all Pareto-optimal plans for all operators.
Hence, it has $O(PK)$-time complexity for $K$ operators to fit on-chip and $P$ plans per operator.
Combined with $\S$\ref{sec:design:algorithm}, the complexity is $O(PK^2N)$ for $N$ operators ($K$ is also the number of possible preload numbers of each operator).

\subsection{Preload Order Permutation}
\label{sec:design:reorder}

\pname{} allows operators to be preloaded in a different order than the execution order.
This has two benefits.

First, reordering helps mitigate interconnect contention.
As the interconnect traffic pressure fluctuates (see $\S$\ref{sec:background:interconnect-util}), the reordering opportunistically reschedules heavy preload traffic to avoid ``rush hours" on the interconnect.

\begin{figure}[t]
    \centering
    \vspace{.1em}
    \includegraphics[width=\linewidth]{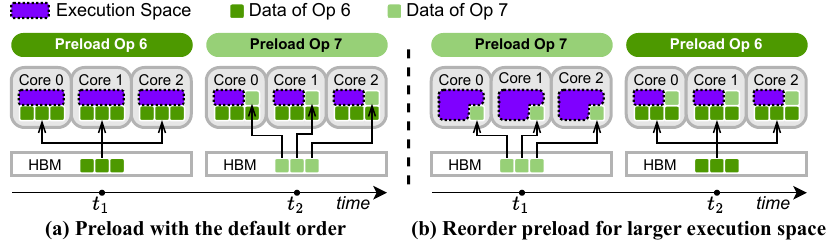} 
    \vspace{-4.5ex}
    \caption{Reorder preloads to allow larger execution space.}
    \label{fig:design:reorder-benefit}
\end{figure}

Second, by reordering the preload of some large operators to a later time, we can save more space for execution by reducing the lifespans of their large memory footprints in the on-chip SRAM.
For instance, in \Cref{fig:design:reorder-benefit}, \textstt{Op6} requires more preload space than \textstt{Op7}.
If we preload in order, the execution space is $1/2$ of the total on-chip memory at time $t_1$.
If we reorder their preloads, the execution space is $5/6$ of the total memory at $t_1$.

As large models consist of thousands of operators, it is unrealistic to test all preload orders (there are $N!$ orders given $N$ operators).
However, most of the orders are invalid, as they overflow the on-chip memory.
If we delay an operator's preload to a late time, its execution will also be delayed.
As operators are executed in order, future operators also cannot execute until this delayed operator completes execution, even if they have already been preloaded into the on-chip memory.
Since there is no free space to preload more operators, and the preloaded operators cannot free their space until executed, the on-chip memory will overflow. 
In practice, \pname{} only needs to explore a reasonable amount of valid preload orders.

\begin{figure}[t]
    \centering
    \vspace{0.3ex}
    \includegraphics[width=\linewidth]{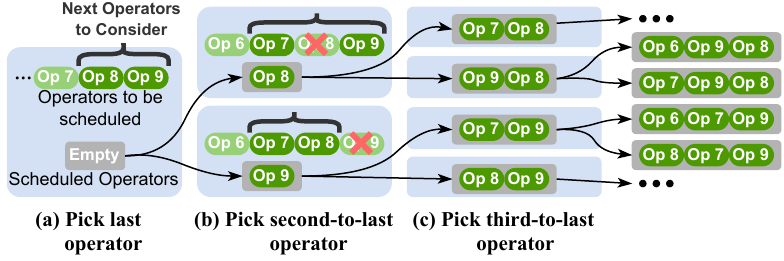} 
    \vspace{-4ex}
    \caption{The generation of candidate preload orders.}
    \label{fig:design:orders}
\end{figure}

\noindent
\textbf{Generate valid preload orders.}
\pname{} enumerates all valid preload orders by scanning through all operators following the inductive operator scheduling order ($\S$\ref{sec:design:algorithm}) and incrementally picking the next operator to preload in each step.

\Cref{fig:design:orders} shows an example of a DL model with 9 operators.
In the first step (\Cref{fig:design:orders} (a)), \pname{} picks the last operator to preload.
We can only fit two operators into the on-chip memory, so either \textstt{Op8} or \textstt{Op9} can be the last operator to preload. This generates two branches for the next step. 

In the second step (\Cref{fig:design:orders} (b)), \pname{} iterates through both branches and picks the second-to-last operator to preload for each branch.
In the upper branch, \textstt{Op8} is already preloaded.
If we choose \textstt{Op6} as the second-to-last operator to preload, both \textstt{Op7} and \textstt{Op9} need to be preloaded before \textstt{Op6}. 
This implies all three operators, \textstt{Op6}, \textstt{Op7}, and \textstt{Op9}, must stay on-chip together because their memory cannot be freed up until \textstt{Op6} is executed.
In our example, as the memory cannot fit all three operators, we can only choose \textstt{Op7} or \textstt{Op9}.
Similarly, in the lower branch, we 
can only choose \textstt{Op7} or \textstt{Op8},
and we do not consider the space requirement of \textstt{Op9} because it can be preloaded after we free up \textstt{Op7} and \textstt{Op8}'s memory.

\pname{} repeats the above process and generates a suffix tree of all valid preload orders.
Given $N$ operators in a model, if we can fit at most $K$ operators on-chip, our search tree has $O(K^N)$ leaves, compared to the original $O(N!)$ search space.

\vspace{0.5em}
\noindent
\textbf{Prune the valid order search space.}
Given the unique characteristics of LLMs, \pname{} can further prune the candidate orders while still being able to find a near-optimal order.

\hlA{First, many operators, such as softmax, preload little or no data from HBM, as they perform in-place computations on the intermediate output.}
For example, OPT-30B~\cite{opt} has 2,269 operators, but 289 of them contribute 99.8\% HBM load volume.
\hlC{Since the remaining 1,980 operators preload little or no data from HBM, reordering their preloads will have negligible performance benefits.
\textit{Thus, \pname{} focuses on reordering only the preloads of operators with high HBM load volume}.
In practice, we only reorder the preload of operators whose tensor sizes are above average (e.g., for LLM decoding, the average size is model size divided by operator count).
For smaller operators that often preload little or no data from HBM, we preload them in order (i.e., \mbox{\textstt{Op i}} will be the \mbox{\textstt{i}}'th preloaded operator).}

Second, an LLM consists of identical transformer layers.
\textit{\pname{} only reorders the preloads within one layer}, and applies the same order to identical layers.
With these rules, \mbox{\pname{}} prunes the search space from $O(K^N)$ to $O(C^H)$.
$H$ is the number of HBM-heavy operators per layer, so $H<<N$ ($H \leq 6$ in most transformer models).
$C$ is the maximum number of HBM-heavy operators per layer that can fit on-chip, so $C<<K$ and $C \leq H$.

For each generated preload order, \mbox{\pname{}} invokes the operator scheduling pass in \mbox{$\S$\ref{sec:design:algorithm}}, forming a $O(C^HPK^2N)$ search space.
\pname{} picks the best end-to-end plan among all preload orders. 
\subsection{Mapping to Hardware}
\label{sec:design:together}

\begin{figure}[t]
    \centering
    \vspace{0.2em}
    \includegraphics[width=\linewidth]{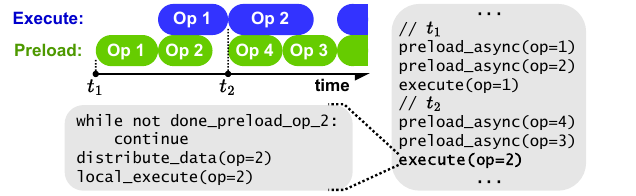} 
    \vspace{-3.5ex}
    \caption{The abstracted device programming model of \pname{}.}
    \label{fig:design:code}
\end{figure}


The execution plan generated by \pname{} specifies all operator's preload order and each operator's partition plans.
\hlcommon{\pname{} maps the plan to an abstracted programming model, which can be applied to generic ICCA chips with off-chip memory.}
As shown in \mbox{\Cref{fig:design:code}}, \pname{} abstracts two key device functions that are generated during compilation. 
(1) \textstt{preload\_async(op=i)} commands all cores to request \textstt{Op}$i$'s data from HBM based on the preload-state partition plan.
(2) \textstt{execute(op=i)} runs \textstt{Op}$i$ on all cores based on the execute-state plan.

\hlB{For the example in \mbox{\Cref{fig:design:code}}, \mbox{\textstt{preload\_async(op=2)}} requests HBM controllers to deliver \mbox{\textstt{Op2}}'s data to each core's SRAM, following the \textit{preload-state plan} (see \mbox{\S\ref{sec:design:allocation}}). When the data delivery completes, the controllers will append a \mbox{\textstt{done\_preload\_op\_2}} tag to the end of the delivered data in each core's SRAM.}

\hlB{Then, \mbox{\textstt{execute(op=2)}} will run in 3 steps when it is called.
First, it waits until \mbox{\textstt{preload\_async(op=2)}} completes, by verifying the value of \mbox{\textstt{done\_preload\_op\_2}} tag in each core's SRAM.}
Second, each core calls \textstt{distribute\_data} to copy shared data from peers, transforming from \textit{preload-state} to \textit{execute-state plan}.
Third, each core calls \textstt{local\_execute} to compute a tile following the \textit{execute-state plan}.

\hlB{To summarize, the hardware enforces three rules for \mbox{\textstt{preload\_async}} and \mbox{\textstt{execute}} calls, using one-way synchronization.}
(1) An \mbox{invocation} of \textstt{execute} blocks all future \textstt{preload\_async}s and \textstt{execute}s, until the invoked \textstt{execute} finishes.
This enforces the operator execution order and specifies which \textstt{preload\_async}s can overlap with an \textstt{execute}.
(2) To enforce the preload order, all \textstt{preload\_async}s execute sequentially.
(3) \textstt{preload\_async(op=i)} does not block any \textstt{execute} except \textstt{execute(op=i)}, as an operator must preload before execution.

%% file: implt.tex
\section{Implementation details}
\label{sec:implt}
\noindent
\textbf{\pname{} compiler framework.}
We implement \mbox{\pname{}} as a generic compiler framework that can support different ICCA chip implementations.
Most \mbox{\pname{}} components are hardware-agnostic.

\noindent
\textul{\textit{(1) \pname{} frontend}} takes a DNN model from ML frameworks like PyTorch~\cite{pytorch-model} as input.
\hlB{The model is first converted into an ONNX graph \mbox{\cite{onnx}}, which represents all operators in the model as a directed acyclic graph.
\pname{} obtains layer information, operator definitions \mbox{\cite{tensor_comprehensions}}, and tensor shapes from the ONNX graph.}
\hlE{\pname{} can support most DL models representable as an ONNX graph.}

\noindent
\textul{\textit{(2) The execution plan generation}} (\S\ref{sec:design:algorithm}--\S\ref{sec:design:reorder}) takes operator partition plans as inputs. \pname{} supports single-operator partition plans generated by compilers that use different parallel execution models~\mbox{\cite{t10,roller,megatron}}.
In our experiments, we use the plans generated with the recent compute-shift execution model proposed in~\cite{t10}, as it represents the state of the art for operator execution on ICCA chips.

For each operator, we enumerate all possible partition plans by representing each plan as a list of integers. For instance, <90,9> evenly slices each dimension of a 2-dimension operator into 90 and 9 parts, forming 90$\times$9=810 tiles.
For each plan, \pname{} decides the mapping of each tile to each core.
It uses different mapping strategies for different network topologies.
\pname{} currently targets ICCA chips with two popular network topologies: all-to-all network and mesh network.
For chip with all-to-all network, \pname{} sequentially maps all tiles, as core locations do not impact the inter-core data transfer cost.
For chip with $N$-dimensional mesh network, \pname{} chooses from plans that partition an operator along at most $N$ dimensions, so it can map each partitioned dimension to a mesh dimension.
Then, \pname{} uses dimension-order routing~\mbox{\cite{tpuv4-optical,waferllm}} to maximize the all-reduce bandwidth.
Besides the two topologies that are used by most ICCA chips today, \pname{} is scalable to support other topologies.

Based on the partition, mapping, and routing information, \pname{}'s cost model estimates each plan's compute, memory, and interconnect costs.
Using the costs of all plans for all operators, \pname{} runs the scheduling, allocation, and reordering procedures in \mbox{\S\ref{sec:design:algorithm}--\S\ref{sec:design:reorder}} to trade-off among performance factors and compose an optimized end-to-end execution plan.
The execution plan generation in \pname{} is implemented in 2.5K lines of code (LoC) of Python.

\noindent
\textul{\textit{(3) The code generation in \pname{}}} generates the kernel code for computing each tile and the inter-core data transfer operations, based on the target hardware and selected partition plans. 
For compute, \pname{} uses code templates from vendor-provided libraries \mbox{\cite{PopLibs}}. 
{For inter-core transfer, \pname{} reserves an 8KB buffer in each core's 624KB SRAM to buffer incoming data, which improves the transfer granularity and performance.}
The code generation in \pname{} is developed in 4K LoC of Python and C++.

\noindent
\textul{\textit{Scalability of \pname{}.}}
\mbox{\pname{}} prunes the search space of a large model to $O(C^HPK^2N)$ complexity.
We list the complexity factors for different models in~\Cref{tab:models}, all using batch size 32 and sequence length 2048.
As model size grows, $N$ scales sub-linearly, while $C$, $H$, $P$, and $K$ change independently.
\hlE{Thus, \pname{}'s search space size scales sub-linearly with the DL model size.}

\begin{figure}[t]
    \centering
    \includegraphics[width=1\linewidth]{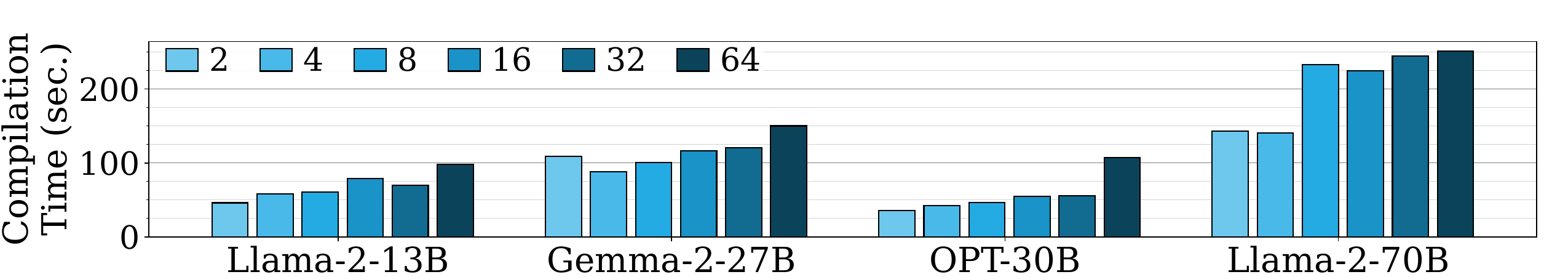} 
    \vspace{-4.5ex}
    \caption{\pname{} compile time for varied model/batch sizes.}
    \label{fig:eval:compile-time}
\end{figure}

\pname{} can generate an end-to-end plan for an LLM on ICCA chip like IPU-POD4 in 5 minutes using a 32-core AMD EPYC 7543 CPU (see \Cref{fig:eval:compile-time}).
On each CPU thread, \pname{} can test a candidate preload order in seconds.
As \pname{} prunes the number of preload orders (e.g., 720 for Llama2-70B), the compilation finishes in minutes.

\renewcommand{\tabcolsep}{4.3pt}
\begin{table}[t]
    \centering
    \caption{
        DL models used in our evaluation. \textit{\underline{C}}: max number of HBM-heavy operators per layer that fit on-chip. \textit{\underline{H}}: number of HBM-heavy operators per layer. \textit{\underline{P}}: max number of plans per operator. \textit{\underline{K}}: max number of operators that fit on-chip. \textit{\underline{N}}: total number of operators.
        We calculate \textit{C} and \textit{K} using the on-chip memory capacity of real IPU-POD4 as an example.
    }
    \vspace{-1ex}
    \footnotesize
    \begin{tabular}{|c|c|c|c|c|c|c|}
    \hline
        \textbf{Name} & \textbf{Description} & \textit{\textbf{C}} & \textit{\textbf{H}} & \textit{\textbf{P}} & \textit{\textbf{K}} & \textit{\textbf{N}} \\\hline
        \specialcell{Llama2-13B~\cite{llama2}} & \specialcell{Large language model (LLM)} & 6 & 6 & 66 & 88 & 1928 \\\hline
        \specialcell{Gemma2-27B~\cite{gemma}} & \specialcell{LLM with Grouped-Query\\Attention (GQA)~\cite{GQA}} & 6 & 6 & 206 & 128 & 2216 \\\hline
        \specialcell{OPT-30B~\cite{opt}} & \specialcell{LLM} & 5 & 6 & 58 & 46 & 2269 \\\hline
        \specialcell{Llama2-70B~\cite{llama2}} & \specialcell{LLM with GQA~\cite{GQA}} & 6 & 6 & 168 & 86 & 3808 \\\hline
        \specialcell{DiT-XL~\cite{dit}} & \specialcell{Diffusion transformer} & 4 & 4 & 123 & 136 & 1521 \\\hline
    \end{tabular}
    \label{tab:models}
    \vspace{-.5ex}
\end{table}

\begin{figure*}
    \centering
    \includegraphics[width=1\linewidth]{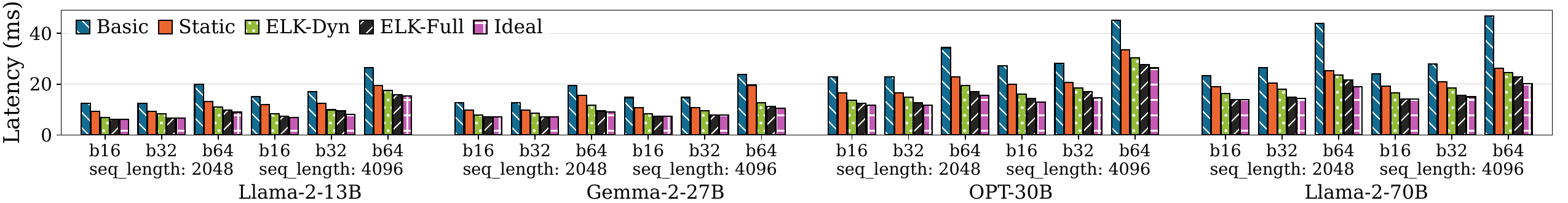} 
    \vspace{-4.5ex}
    \caption{The per-token serving latency of various models and batch sizes on 4 ICCA chips with 16TB/s HBM.}
    \label{fig:eval:e2e}
\end{figure*}

\noindent
\textbf{Emulation framework.}
As the ICCA chip we can access (IPU-POD4) does not have HBM, we build an emulation framework using a real IPU-POD4.
\hlC{The pod has 4 IPU MK2 chips with a total of 5,888 cores, 3.5GB on-chip memory, and 640GB/s inter-chip bandwidth.}
\hlB{By default, we use model parallelism~\mbox{\cite{megatron}} across the four chips, since it incurs little inter-chip communication overhead, because the activation tensor to be reduced across chips is usually small.}
To obtain HBM access latencies, our framework uses an acknowledged memory simulator~\mbox{\cite{dramsim3}}.
We evenly slice each tensor across all HBM modules to balance traffic, and sequentially place tensors in HBM.
{The HBM can easily saturate its bandwidth when \mbox{\pname{}} sequentially reads data at tensor granularity (tensor sizes range from 43 to 219 MB).}
Based on the tensor placement, we generate memory traces of all tensors to obtain HBM latencies from the memory simulator.

\hlcommon{The framework then executes the end-to-end plan generated by \pname{} on IPU-POD4, where it computes each tile on each core and moves shared data between cores based on the partition plans selected by \pname{}.}
To emulate HBM accesses, one core acts as an HBM controller to broadcast ``HBM data'' to other cores and apply HBM latencies by delaying each broadcast.
\hlB{To synchronize execution with preload, the ``controller'' core also appends \mbox{\textstt{done\_preload\_op\_i}} tags (i.e., arbitrary constants, see \mbox{\S\ref{sec:design:together}}) to the end of broadcasted data, allowing receiver cores to check whether a preload has finished.} 

As \pname{} homogeneously partitions each tensor to cores following common operator tiling strategies~\cite{roller,tvm,xla}, all cores receive tensor tiles of the same size during each preload. 
Thus, we emulate the interconnect traffic caused by preload by using one ``controller'' core to broadcast data to all cores.
The broadcast saturates the interconnect and the inbound links on receiver cores,
emulating the contention between inter-core data sharing and operator preload.

\noindent
\textbf{Simulation framework.}
To conduct sensitivity analysis and design space exploration, we build an event-driven simulator for ICCA chips, which simulates all cores, network links, and off-chip HBM accesses.
For each core, we simulate a local SRAM, a compute pipeline, and a network agent that sends/receives data to/from other cores.
For each network link, we model its latency and bandwidth~\cite{ahead:ipdpsw19,loggp:spaa95}.
Based on the execution plan generated by \pname{}, we derive the simulation events at tile granularity, including computing a tile on a core, transferring a tile over a specific network link, and fetching a tile from the off-chip HBM.
Each core/link maintains its event queue to execute its events sequentially.
For an all-to-all network, we model HBM controllers as dedicated nodes in the network (see $\S$\ref{sec:background:ipu-arch}).
For a mesh network, we attach HBM controllers to the edges of the mesh grid.
{To simulate a multi-chip system, we track the in-flight inter-chip transfer events and cap their total bandwidth.
We also use our real IPU-based emulator to validate our simulator.}



%% file: eval.tex
\section{Evaluation}
\label{sec:eval}

With our emulation framework, we show that on average, \hlA{\pname{} achieves (1) \textbf{94.84\%} of the performance of an ideal roofline design (\mbox{\S\ref{sec:eval:e2e}}),} (2) \textbf{89.52\%} inter-core interconnect bandwidth utilization, and almost ideal HBM and FLOPS utilization relative to the roofline (\S\ref{sec:eval:util}).
With our simulator, (3) we demonstrate \pname{} enables design space exploration for scaling compute, communication, and off-chip memory accesses for ICCA chips. We report our insights in \S\ref{sec:eval:sens}. 


\vspace{-0.2em}
\subsection{Experimental Setup}
\label{sec:eval:setup}



\noindent
\textbf{Workloads.}
We examine the inference decoding phase of differently sized LLMs (see \Cref{tab:models}), using varied batch sizes and sequence lengths.
We also test a stable diffusion model (see Figure \mbox{\ref{fig:eval:vary-cores}}) and LLM training (see \Cref{fig:train}).


\noindent
\textbf{Emulator setup.}
We emulate 4 HBM3E modules~\cite{hbm3e} per ICCA chip, following a state-of-the-art (SOTA) GPU~\cite{blackwell}.
With 4 ICCA chips, we have 16TB/s total HBM bandwidth.

\noindent
\textbf{Simulator setup.}
We simulate 4 chips and 16TB/s HBM bandwidth by default.
{The configuration (compute and local SRAM) of each core and the latency/bandwidth of each network link are the same as the emulator setup by default,
and packets that share one NoC link are scheduled sequentially.}
We simulate both all-to-all and 2D mesh networks.
For all-to-all network, we follow the IPU-POD4 architecture \mbox{\cite{ipu2}}.
For mesh network, each core can simultaneously communicate with all its neighbors (up to 4 in a 2D mesh)~\mbox{\cite{tenstorrent-TT-METALIUM}}.



\noindent
\textbf{Baselines.}
As there is no open-sourced compiler for ICCA chip with HBM,
we conduct an ablation study by creating two baselines that extend SOTA compilers for ICCA chips~\cite{t10} to support HBM, and an \pname{} variant that disables preload reordering ($\S$\ref{sec:design:reorder}).
We also compare \pname{} to an ideal roofline.
In brief, we compare these designs:

\begin{itemize}[leftmargin=*]

\item \textbf{\naive{}}:
The design follows existing DL compilers to optimize on-chip execution. It maximizes the execution space and uses the remaining space to preload the next operator.

\item \textbf{\baseline{}}:
Following the SOTA compiler T10~\cite{t10} developed for ICCA chips, we extend it to jointly optimize on-chip execution and off-chip loading. 
First, it follows SambaNova~\mbox{\cite{samba-sn40}} to preload multiple operators in advance, by reserving a preload space.
Then, it find the fastest execution plan for each operator given the remaining execution space size.
We further improve the design by finding the best static preload and execution space sizes for the entire DL model
(the sizes will not change throughout the model execution).
When preloading a set of operators, all operators use either the preload-state plan with the largest memory footprint or the plan with the smallest footprint, whichever is faster.


\item \textbf{\textit{\pname{}-Dynamic (\fixorder{})}}: 
A partial design of \pname{}, which optimizes the preload-execution overlap (\S\ref{sec:design:algorithm}) and on-chip memory allocation (\S\ref{sec:design:allocation}).
\hlD{This design represents \pname{}'s performance \textit{without} preload order permutation (\mbox{\S\ref{sec:design:reorder}}).}

\item \textbf{\full{}}: The full \pname{} design, which enables all optimizations, including the preload order permutation (\S\ref{sec:design:reorder}).

\item \textbf{\ideal{}}: 
The theoretical roofline performance, where each of preload and execution has its own interconnect (i.e., no interconnect contention) and full-sized on-chip memory (i.e., no memory space contention).
Each operator uses the minimum preload space to emulate the benefits of maximum preload number,
and the data distribution phase has zero latency to emulate the benefits of maximum preload space per operator.


\end{itemize}

\subsection{End-to-end Performance}
\label{sec:eval:e2e}

\Cref{fig:eval:e2e} shows the per-token generation latency of LLM decoding on our emulator.  
On average, \mbox{\full{}} outperforms \mbox{\naive{}} by \textbf{1.87$\times$} (up to \textbf{1.93$\times$}), \mbox{\baseline{}} by \textbf{1.37$\times$} (up to \textbf{1.49$\times$}), and achieves \textbf{94.84\%} of the ideal performance.
The performance of \pname{} also scales well with increasing batch size and sequence length. 
Notably, Gemma2-27B and Llama2-70B can achieve latencies similar to those of smaller LLMs, since they use Grouped-Query Attention~\cite{GQA}.

\begin{figure} 
\vspace{0.5ex}
    \centering
  \subfloat[\footnotesize{Breakdown of per-token latency.}\label{fig:eval:e2e-breakdown-a}]{%
       \includegraphics[width=0.49\linewidth]{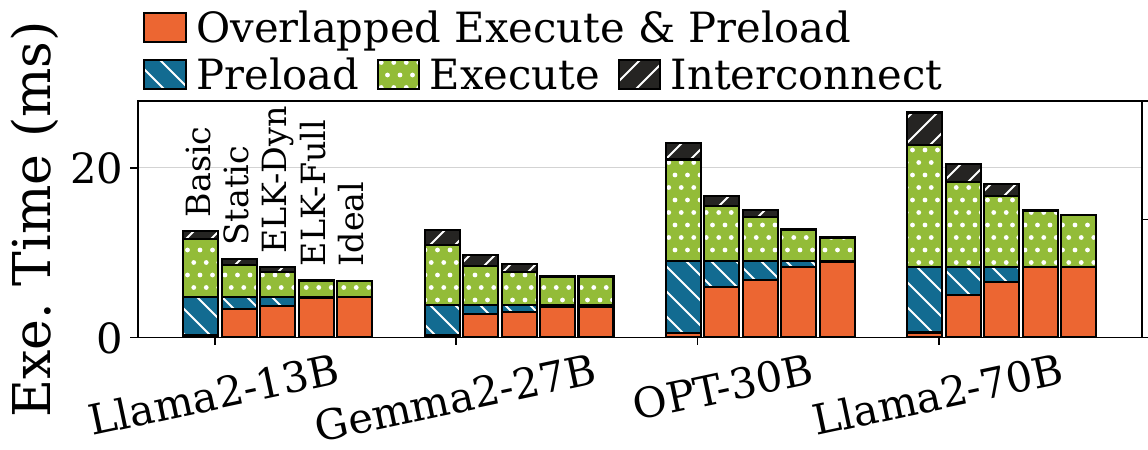}\vspace{-0.5em}}
    \hfill
  \subfloat[\footnotesize{Avg. HBM bandwidth utilization.}\label{fig:eval:e2e-breakdown-b}]{%
        \includegraphics[width=0.49\linewidth]{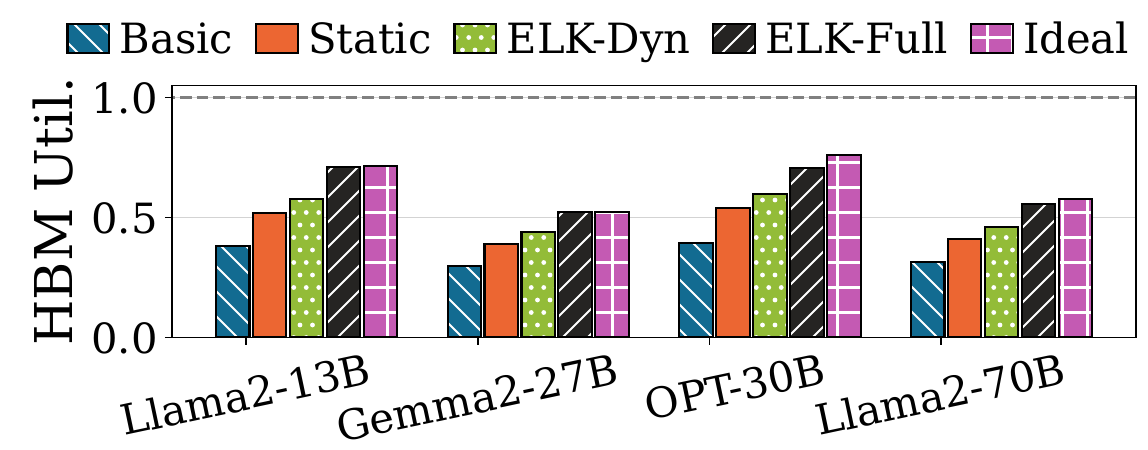}\vspace{-0.5em}}
    \\
    \vspace{1ex}
  \subfloat[\footnotesize{Average interconnect utilization. Top column parts are inter-core data sharing; bottom are operator preload.}\label{fig:eval:e2e-breakdown-c}]{%
        \includegraphics[width=0.49\linewidth]{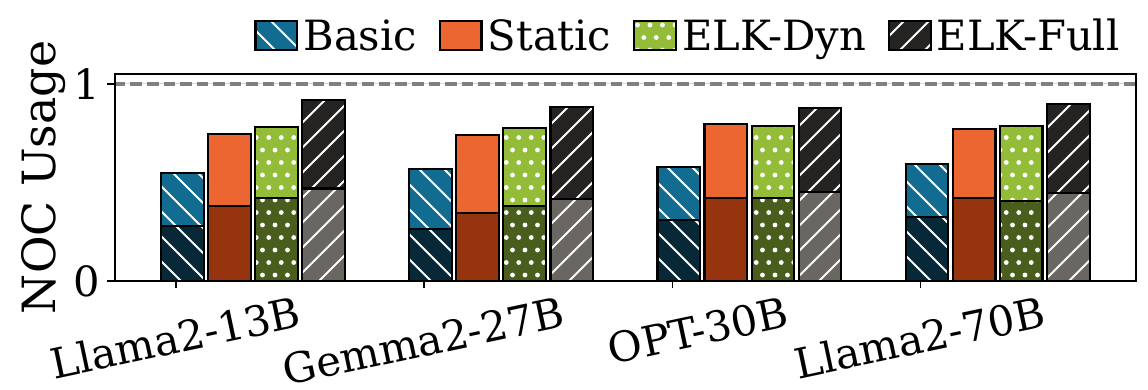}\vspace{-0.5em}}
    \hfill
  \subfloat[\footnotesize{Average TFLOPS throughout the execution of each model.}\label{fig:eval:e2e-breakdown-d}]{%
        \includegraphics[width=0.488\linewidth]{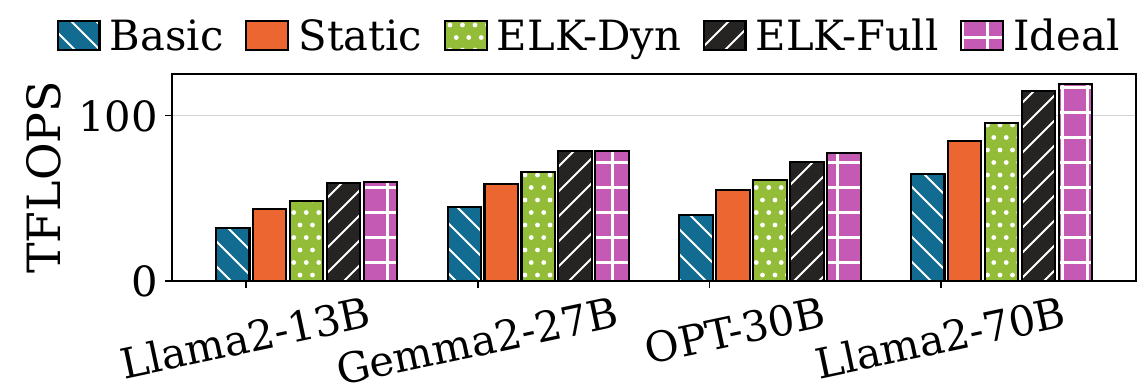}\vspace{-0.5em}}
\vspace{-1.3ex}
  \caption{Execution breakdown and resource utilization. In (a), we categorize total time into preload (HBM is busy), execute (cores are busy), overlapped execute/preload, and interconnect (execute/preload stopped by busy interconnect).}
  \label{fig:eval:e2e-breakdown}
\end{figure}

\begin{figure}[t]
    \centering
    \includegraphics[width=\linewidth]{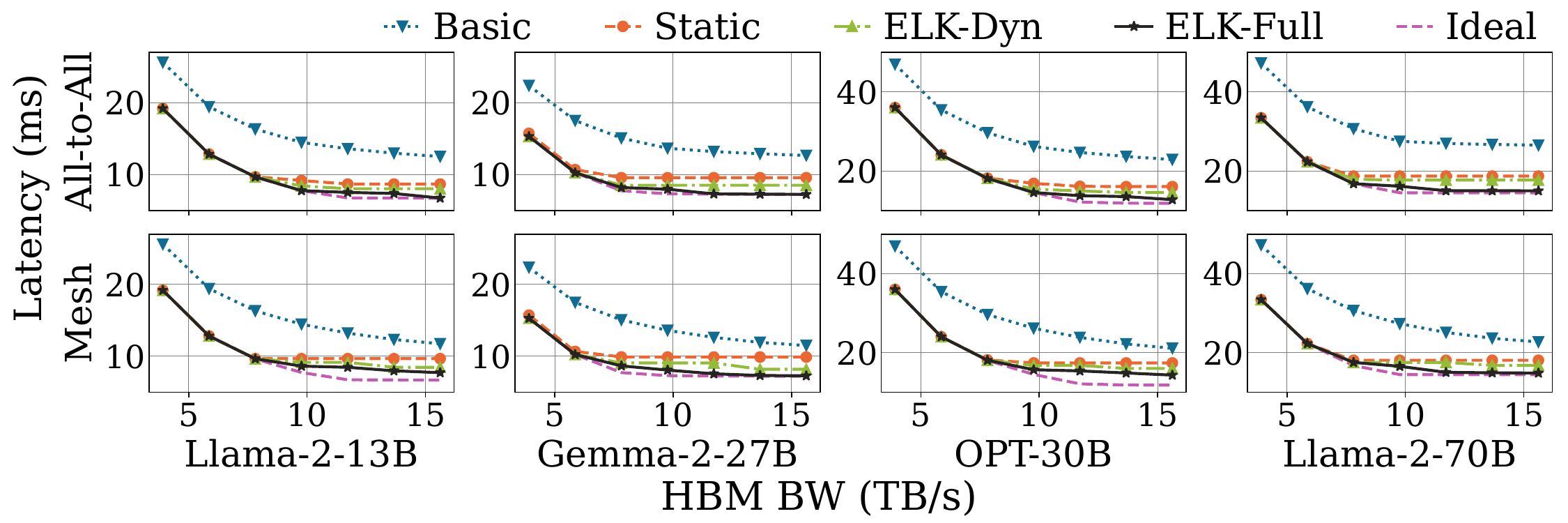} 
    \vspace{-4.5ex}
    \caption{Per-token latency at varied HBM bandwidths.}
    \label{fig:eval:vary-bw}
\end{figure}

\noindent
\textbf{Inference latency breakdown.}
In \mbox{\Cref{fig:eval:e2e-breakdown}} (a), we break total time into four categories: (1) \textit{preload} (HBM is loading), (2) \textit{execute} (cores are computing/sending data), (3) \textit{overlapped preload \& execute}, and (4) \textit{interconnect} (HBM/cores are stalled by interconnect contention).
We only show batch size 32 and sequence length 2048 due to space limits.
%
%
\naive{} always poorly overlaps preload and per-core execution.
By preloading more operators, \baseline{} increases the overlap time by \textbf{11.26$\times$}, but is limited by fixed preload and execution space sizes.
\fixorder{} overlaps better by adjusting the on-chip memory allocation based on operators' demands, but suffers from interconnect congestion and misses preload opportunities (when the available preload space is too small for the next operator, but can fit a future operator).
By reordering preloads with an average edit distance of \textbf{2.9} steps, \full{}
eliminates \textbf{87.65\%} of interconnect congestion overhead over \mbox{\fixorder{}}.
\full{} also reduces the non-overlapped preload time to \textbf{0.037\%} of the total, because of reduced on-chip memory contention. 

\subsection{Hardware Resource Utilization}
\label{sec:eval:util}

\noindent
\textbf{HBM bandwidth.}
\Cref{fig:eval:e2e-breakdown} (b) shows the average HBM bandwidth utilization for each design.
\naive{} uses \textbf{34.7\%} of the bandwidth. It only preloads the next operator, causing HBM idleness.
\baseline{} utilizes \textbf{46.42\%} by preloading multiple operators in advance, 
but the fixed-size preload space limits the preload opportunity and fails to keep HBM busy.
\fixorder{} achieves \textbf{51.97\%} utilization by allowing larger preload spaces. 
\full{} further achieves \textbf{62.40\%} utilization with preload reordering, which is close to the \textbf{64.38\%} utilization of \ideal{}.
Note that \ideal{} does not fully utilize HBM bandwidth, as there is more bandwidth available than necessary to load the entire model during execution.


\noindent
\textbf{Interconnect bandwidth.}
\Cref{fig:eval:e2e-breakdown} (c) shows the interconnect bandwidth utilization for each design.
\naive{} only utilizes \textbf{57.25\%} of the bandwidth.
\baseline{} and \fixorder{} can better overlap execute and preload, but their utilizations are still only \textbf{76.33\%} and \textbf{78.28\%}.
\full{} achieves \textbf{89.52\%} utilization, since preloads with low interconnect traffic can be reordered to match operator execution periods with high traffic. This alleviates interconnect contention.
We cannot make a fair comparison with \ideal{}, because \ideal{} is modeled using two separate interconnects for preload and execute.

\noindent
\textbf{FLOPS.}
In \Cref{fig:eval:e2e-breakdown} (d), 
\full{} achieves \textbf{81.06} TFLOPS. 
Though our emulator theoretically offers 1000 TFLOPS for MatMuls or 31.2 TFLOPS for other operations, LLM inference is bandwidth-bound, and actual TFLOPS is limited by on-chip data transfer (the interconnect utilization is already as high as 90\%). \full{}'s TFLOPS is already close to that of \ideal{}.


\subsection{Design Space Exploration for ICCA Chips}
\label{sec:eval:sens}

To understand how to scale future ICCA chips, we use our ICCA chip simulator (\S\ref{sec:implt}) to explore the performance impacts of different network topologies, interconnect bandwidths, HBM bandwidths, and compute capabilities (FLOPS).

\noindent
\textbf{(1) Higher HBM bandwidth improves the per-token latency, but the benefit will diminish due to higher interconnect contention.}
In \Cref{fig:eval:vary-bw}, we examine \pname{} with various HBM bandwidths and interconnect topologies.
When HBM bandwidth is low, all designs are bounded by HBM.
With more HBM bandwidth (e.g., $\approx$8TB/s for Llama2-70B), the performance becomes bounded by the interconnect and per-core execution.
Also, since mesh-based network takes multiple hops to deliver HBM data to cores, it suffers higher interconnect contention than all-to-all network.
Thus, it is harder for \full{} to match with \ideal{} on mesh, especially for non-GQA models like Llama2-13B and OPT-30B, as they fetch more KV cache data from HBM.

\begin{figure}[t]
    \centering
    \includegraphics[width=\linewidth]{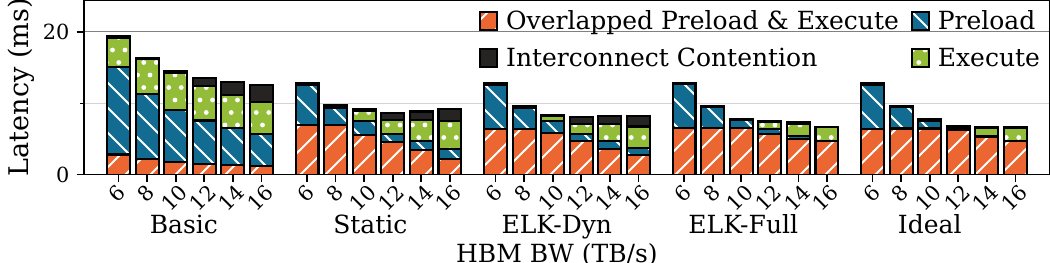} 
    \vspace{-3.5ex}
    \caption{Breakdown of LLama2-13B per-token latency with varied HBM bandwidths on all-to-all network. {We categorize total time into preload (HBM is loading), execute (cores are computing or sending data), overlapped preload/execute, and interconnect contention (preload/execute stopped by busy interconnect).} We only show one case due to space limits.}
    \label{fig:eval:breakdown}
\end{figure}

In \Cref{fig:eval:breakdown}, we show the latency breakdown of the interconnect contention.
For \naive{}/\baseline{}/\fixorder{}, contention increases with higher HBM bandwidth, as faster HBM needs more interconnect bandwidth to deliver data to cores.
%
\full{}'s reordering allows more preload opportunities which better utilize the faster HBM to eliminate the contention.


\begin{figure}[t]
    \centering
    \includegraphics[width=\linewidth]{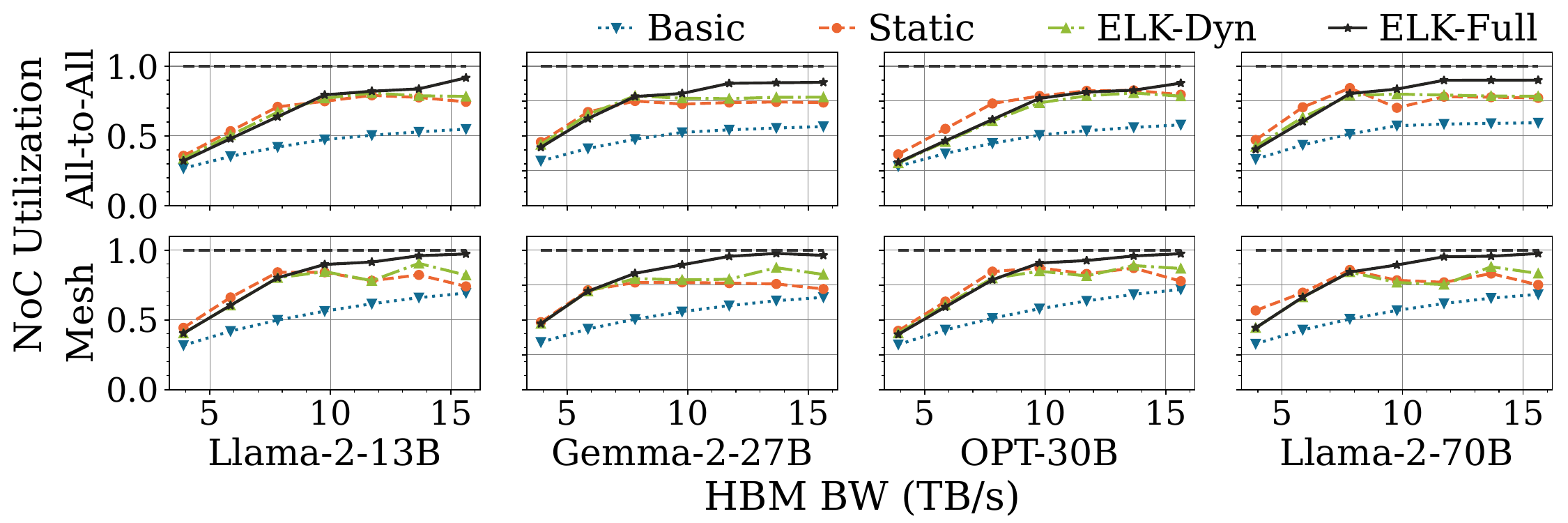} 
    \vspace{-5ex}
    \caption{Interconnect utilization at varied \footnotesize HBM \small bandwidths.}
    \label{fig:eval:noc-util}
    \vspace{-.5ex}
\end{figure}

In \Cref{fig:eval:noc-util}, we compare the interconnect utilization between the all-to-all and mesh topologies.
While achieving similar serving latencies, mesh chips always experience higher interconnect utilization than all-to-all, since mesh takes multiple hops to deliver HBM data to cores.
For both topologies, \full{} is the only design that can almost fully utilize the interconnect.
In other designs, HBM data delivery often occupies the interconnect and stalls the execution.

\begin{figure}[t]
    \centering
    \includegraphics[width=\linewidth]{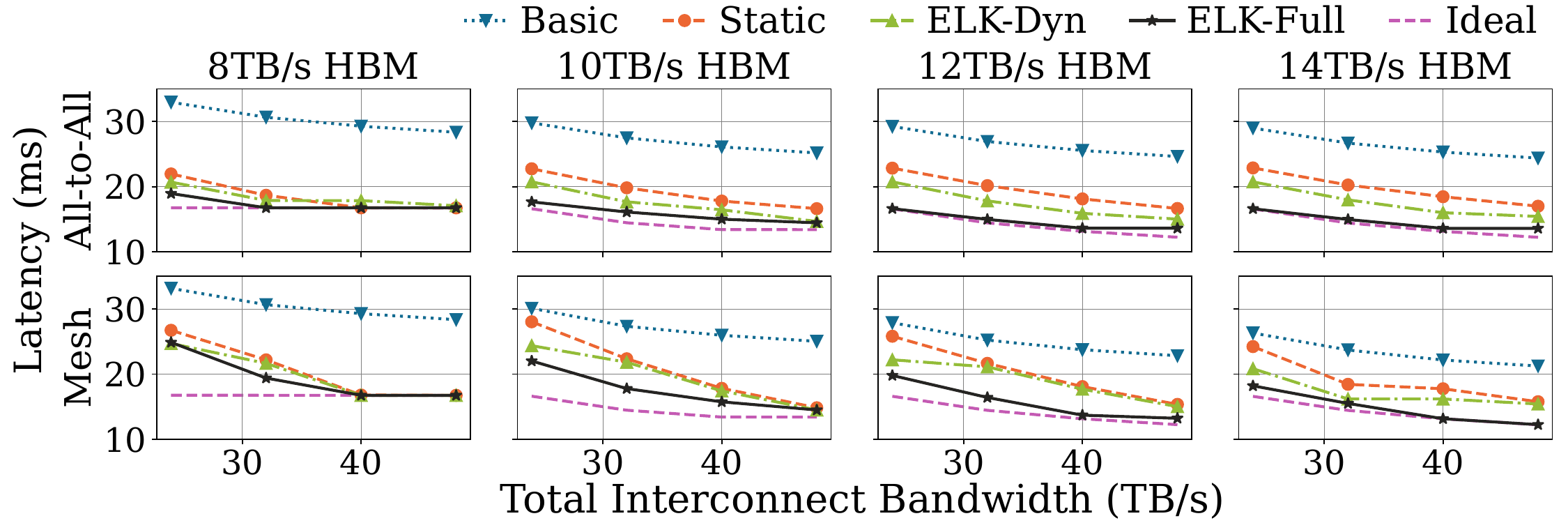} 
    \vspace{-4.5ex}
    \caption{Llama2-70B latency of at varied NoC bandwidths.}
    \label{fig:eval:noc-hbm-bw}
    \vspace{-.5ex}
\end{figure}

\noindent
\textbf{(2) The interconnect and HBM bandwidths should scale together to avoid performance bottlenecks.}
In \Cref{fig:eval:noc-hbm-bw}, we examine how the interconnect bandwidth impacts the performance under different HBM bandwidths.
When the HBM bandwidth is low (e.g., 8TB/s per 4 chips), increasing the interconnect bandwidth beyond a certain point (e.g., 40TB/s) has no benefit, since HBM is the bottleneck.
With higher HBM bandwidth, performance scales with the interconnect bandwidth, and \full{} can best utilize both bandwidths to achieve near-\ideal{} performance.
Compared with all-to-all, the performance of mesh is more sensitive to the interconnect bandwidth.
This matches the finding that mesh-based ICCA chips utilize the interconnect more heavily (\Cref{fig:eval:noc-util}).

\begin{figure}[t]
    \centering
    \includegraphics[width=\linewidth]{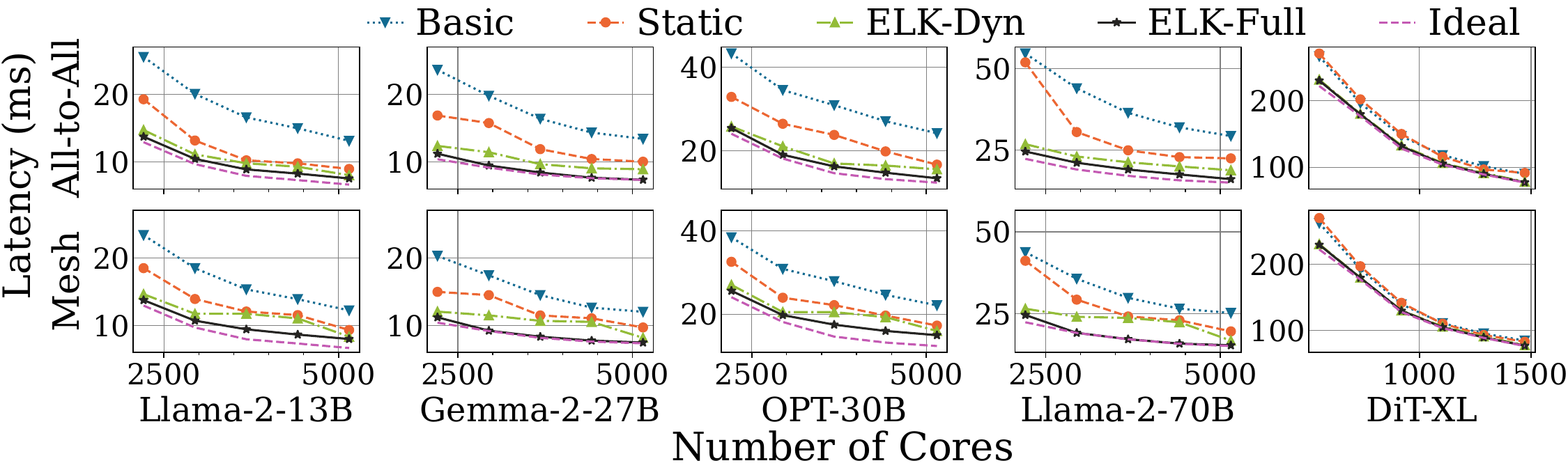} 
    \vspace{-4.5ex}
    \caption{Per-token latency at varied core counts.}
    \label{fig:eval:vary-cores}
\end{figure}

\noindent
\textbf{(3) \pname{} enables scalable performance for ML inference workloads as we scale the ICCA chip.}
In \Cref{fig:eval:vary-cores}, we change the number of cores while setting the HBM bandwidth to 2.7GBps/core to match prior setups.
\mbox{\full{}} significantly outperforms other designs regardless of core counts.
\mbox{\full{}} reduces the average latency by \textbf{1.71$\times$} over \mbox{\naive{}} and \textbf{1.36$\times$} over \mbox{\baseline{}}.
We also examine DiT-XL, a state-of-the-art stable diffusion model, on one ICCA chip (up to 1472 cores).
\mbox{\full{}}'s benefit on DiT-XL is less obvious than on LLMs, since DiT-XL is compute-intensive and less affected by preload efficiency.
However, \mbox{\full{}} still outperforms other designs on DiT-XL and achieves near-ideal performance.

\begin{figure} 
    \centering
  \subfloat[\footnotesize{All-to-all interconnect.}\label{fig:mesh-2080}]{%
       \includegraphics[width=0.48\linewidth]{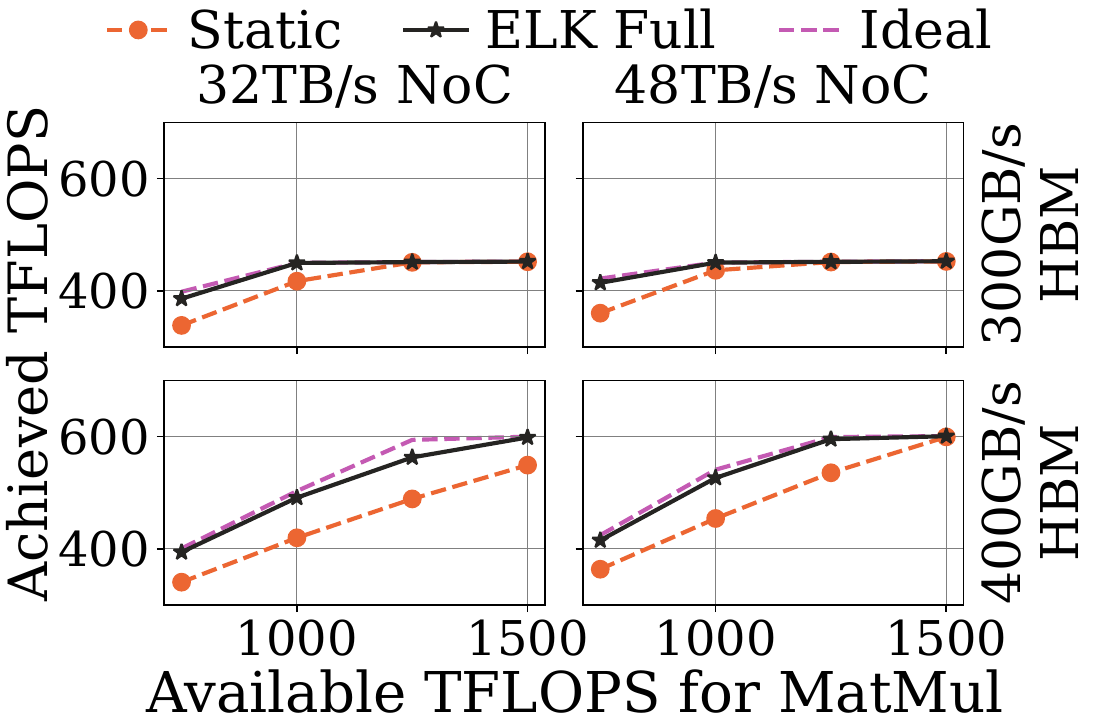}\vspace{-0.2em}}
    \hfill
  \subfloat[\footnotesize{Mesh interconnect.
  }\label{fig:mesh-4160}]{%
        \includegraphics[width=0.48\linewidth]{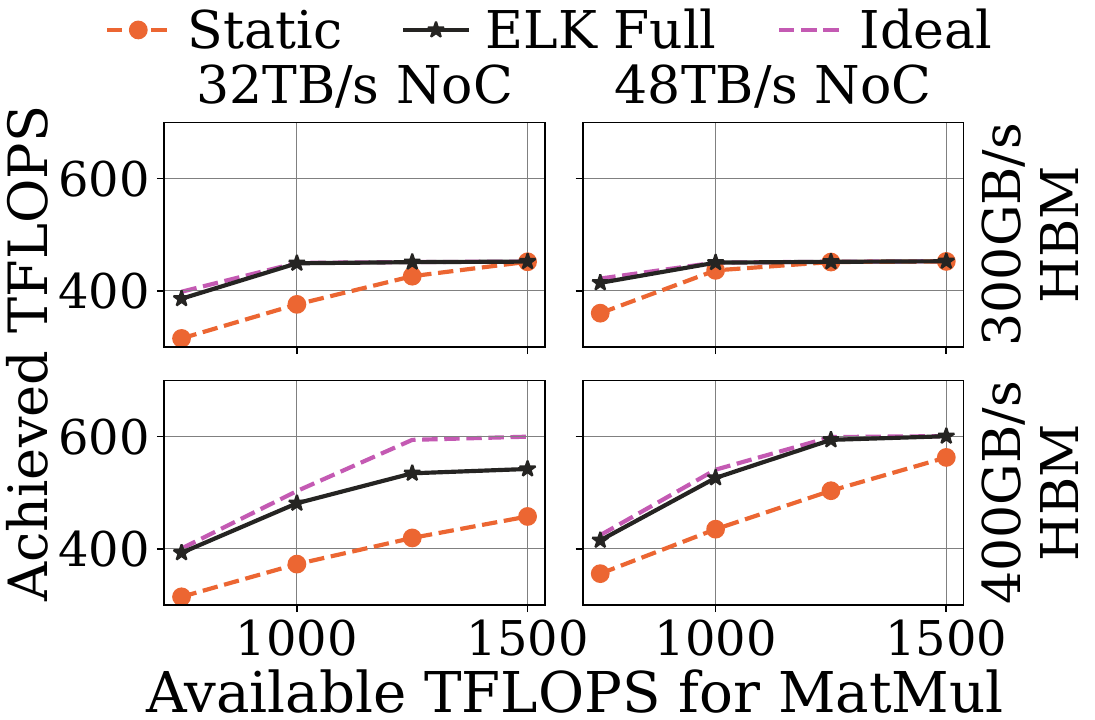}\vspace{-0.2em}}
\vspace{-1.5ex}
  \caption{Average TFLOPS during the training of Llama2-13B, given varied amount of computation resources.}
  \label{fig:train}
\end{figure}

\noindent
\textbf{(4) ICCA chips can also benefit ML training by properly tuning the compute, communication, and off-chip memory access.}
In \Cref{fig:train}, we examine the forward pass of training Llama2-13B with varied available FLOPS and interconnect/HBM bandwidths (the backward pass has similar trends).
Unlike decoding, training is compute-intensive, scaling only interconnect/HBM bandwidth has little impact.
With 400GB/s HBM bandwidth, it is sufficient to fulfill more than 600 TFLOPS.
\hlF{Thus, for compute-intensive workloads, the ICCA chips should focus on scaling the FLOPS, and can therefore be paired with cheaper memory (e.g., GDDR/LPDDR/DDR) to reduce manufacturing costs.}
Note that the achieved FLOPS is often lower than the peak FLOPS of the hardware, because only MatMul operators with perfect shapes can fully utilize the FLOPS of specialized tensor cores.



%% file: discussion.tex
\section{Discussion and Future Work}
\label{discussion}


\noindent
\textbf{Apply \pname{} to GPUs.}
The latest NVIDIA GPU also uses inter-core links to connect its stream multiprocessors (SMs)~\cite{hopper}.
It groups SMs into clusters. SMs in the same cluster are connected via direct inter-SM links, while different clusters can only exchange data via the global L2 cache.
On current GPUs like H100~\mbox{\cite{h100-hbm}}, the aggregated inter-SM bandwidth is close to the HBM bandwidth, so it will suffer from significant interconnect contention (\ding{173} in Figure~\mbox{\ref{fig:background:contention}}). As future work, we wish to extend \mbox{\pname{}} to GPU and investigate the design space for optimizing GPU's interconnect architecture.

\vspace{0.2em}
\noindent
\textbf{Apply \pname{} to MoE.}
{\pname{} can support dynamic mixture-of-experts (MoE) models.
In MoE, an operator may choose different parameter tensors (i.e., experts) based on the input token.
At compile time, as all experts have the same shape, \pname{} will optimize the execution plan based on a generic expert.
\pname{} will schedule the preload of an expert to a time after the model selects which expert to use (e.g., after the expert routing operator or the expert prediction~\mbox{\cite{predict-is-all-moe-need}}).
On execution, the chip preloads expert tensors using the partition plans given by \pname{} and the expert indices selected at runtime.}

\vspace{0.2em}
\noindent
\hlD{\textbf{Apply \pname{} to other optimization objectives.}
While \pname{} currently optimizes the performance, it can be adapted to support optimizing for a wide variety of objectives, by replacing the performance-based cost model in \mbox{\S\ref{sec:design:allocation}} with others (e.g., optimize power by adapting a cost model that estimates power usage).}

\vspace{0.2em}
\noindent
\hlcommon{\textbf{Apply \pname{} to other execution models.}
For different ICCA chip implementations, 
they may have different execution models \mbox{\cite{plasticine:isca2017, t10, tenstorrent-TT-METALIUM}}.
For example, SambaNova chips \mbox{\cite{samba-sn40}} support a spatial pipeline execution model that runs different operators on different sets of cores \mbox{\cite{plasticine:isca2017}}.
The pipelined execution keeps model weights stationary on each core and lets activation tensors flow through cores.
This enables significantly higher serving throughput, though the latency of each serving request may increase if there are too many pipeline stages.
This execution model also experiences the resource constraints in \mbox{\S\ref{sec:background:challenges}}.
Specifically, it still needs to use HBM to swap the model weights inside each core's SRAM, unless it uses the SRAM of hundreds of chips to store an entire LLM.
Thus, the pipelined execution has to (1) reserve SRAM for both currently executing data and newly preloaded data and (2) use the interconnect for both inter-core data transfer and HBM data loading, requiring the compiler to consider the resource constraints in \mbox{\S\ref{sec:background:challenges}}.
To optimize for this spatial pipeline execution model, we can modify \pname{}'s search algorithm to explore the new scheduling space of this model (e.g., decide the number of pipeline stages per chip and the number of cores per stage).
We wish to explore the optimization space of various execution models as future work.}

%% file: related.tex
\section{Related Work}
\label{sec:related}



\noindent
\textbf{Deep learning compilers.}
Many DL compilers~\cite{tvm, roller, korch, GraphTurbo:osdi2023, soma, g10} were designed for architectures without inter-core links (e.g., GPU and TPU).
A few compilers serve models purely on-chip, they did not consider off-chip memory~\cite{wse2,t10,groq-llm}. 
As prior compilers~\cite{tvm,tensorir,ansor,roller,t10}
focused on the optimization of tile partitioning of a single operator, \pname{} can utilize them to generate each operator's partition plans.
\hlF{\pname{} is also compatible with other optimization techniques like quantization \mbox{\cite{ladder}} and alignment with memory bank \mbox{\cite{roller}}, as they do not change the execution pattern.}


\vspace{0.1em}
\noindent
\hlE{\textbf{ML optimizations with operator fusion.}}
DL frameworks~\cite{xla, astitch, welder:osdi23, soma, flashatt, clusterfusion} improve on-chip data reuse by fusing multiple operators.
{For example, SoMa \mbox{\cite{soma}} explores fusion opportunities to reuse intermediate tensors between operators in a global on-chip buffer with limited size.
As ICCA chips have large distributed SRAM (e.g., up to 900MB per chip) that can buffer an entire intermediate tensor, they can reuse it between operators without fusion.
However, ICCA chips face unique challenges in distributed SRAM allocation, inter-core interconnect contention, and their impacts on HBM preload.
Thus, \pname{} focuses on the interplay of these performance challenges.}
{For ICCA chips with less SRAM, \pname{} can still support fusion by treating each fused operator as one operator.
For example, we can fuse two consecutive MatMul operators into one operator~\mbox{\cite{flashatt}}, which is treated as one operator with three input tensors in \pname{}.}

\vspace{0.1em}
\noindent
\textbf{Compilers for ICCA chips.}
Dataflow compilers optimize DL execution on interconnected cores by mapping operators to a pipeline~\cite{plasticine:isca2017, amos:isca2022, inter-layer, flexflow-dataflow, sara, Scalable_interconnects, waferllm}.
While they optimize on-chip execution and communication, they use off-chip HBM differently. 
SambaNova SN40L \mbox{\cite{samba-sn40}} swaps parameters between HBM and DDR to serve different expert models. Tenstorrent \mbox{\cite{tenstorrent-TT-METALIUM}} optimizes intra-operator tile size to reduce the total off-chip memory access volume.
T10 \mbox{\cite{t10}} optimizes on-chip execution without considering off-chip memory,
so it cannot support LLMs that exceed on-chip capacity (e.g., $<$1GB per chip).
\hlB{Specifically, T10 can only optimize the placement of a fixed set of operators in a fixed amount of SRAM.
However, as an ICCA chip loads new operators from the HBM and frees old ones from the SRAM, the set of operators in the SRAM changes dynamically.
In addition, the HBM accesses will interfere with the on-chip execution due to the SRAM and NoC contentions.
As T10 cannot consider all performance tradeoffs together, even after we extend T10 to support HBM, the extended design (i.e., \baseline{} in \mbox{\S\ref{sec:eval:setup}}) still performs poorly on ICCA chips with HBM.}
To maximize performance for generic ICCA chips, \mbox{\pname{}} considers the interplay among off-chip HBM, on-chip SRAM, inter-core interconnect, and per-core execution on top of the on-chip execution optimizations enabled by existing compilers for ICCA chips.





\vspace{0.1em}
\noindent
\textbf{Distributed model execution.}
To run DL models on distributed nodes, prior works overlap computation with inter-device communication.
They optimize inter-chip collective communication~\cite{Centauri, themis, coconet, TPU-comp-comm}, device grouping on various network topologies~\cite{alpa, metis:atc24, amp:nips22, Galvatron}, and workload collocation in device clusters~\cite{v10, neu10, topo_aware_virt, neu10_hotos}.
\mbox{\pname{}} targets intra-chip optimization and faces other challenges.
{Besides compute and communication, {\pname{}} also considers HBM data accesses and their impacts on the usages of per-core SRAM and inter-core links.}
\hlB{\pname{} is compatible with various distributed execution frameworks and different parallelism types when using multiple ICCA chips,
where \pname{} optimizes the execution of each ICCA chip.}

\vspace{0.1em}
\noindent
\hlD{\textbf{ML- and SAT-based compiler optimizations.}
Existing studies like Autocomp \mbox{\cite{llm-aided-tensor-optim}} use LLMs to optimize kernel code for tensor accelerators, demonstrating significant performance advantages over vendor libraries.
The code generation in \pname{} (\mbox{\S\ref{sec:implt}}) can leverage these works to further optimize the per-core tile computation, by replacing the code templates from vendor libraries with the kernels optimized by LLMs.
\pname{}'s scheduling algorithm (\mbox{\S\ref{sec:design:algorithm}--\S\ref{sec:design:reorder}}) does not rely on LLMs, since it can already find execution plans with near-\ideal{} performance (see \mbox{\S\ref{sec:eval:e2e}}) in short compile time (see \mbox{\Cref{fig:eval:compile-time}}) using the host CPU.
Both \pname{}'s comprehensive optimization space and its pruning techniques based on hardware information (e.g., SRAM size, core count) contribute to its success.}
\hlD{Other prior works \mbox{\cite{sat-logic-constrain,scheduling-and-sat}} use SAT solvers to solve scheduling problems.
However, it is inefficient to use SAT solvers in \pname{}, because their runtime grows exponentially with the number of boolean variables to solve \mbox{\cite{gong2017survey}}.
For example, to solve \pname{}'s allocation problem in \mbox{\S\ref{sec:design:allocation}} using SAT, we need to assign one boolean variable to each possible execution plan of each operator to indicate whether this plan is selected.
Thus, the solver's runtime grows exponentially with the total number of possible plans of all operators.
In comparison, the complexity of our allocation algorithm in \mbox{\S\ref{sec:design:allocation}} grows linearly with the total number of plans (i.e., $O(PK)$ for $K$ operators that each has $P$ possible plans).}

%% file: conclusion.tex
\vspace{0.1em}
\section{Conclusion}
\label{sec:conclusion}
\vspace{0.1em}

We study the performance trade-offs of generic ICCA chips that support off-chip memory, and develop
a DL compiler framework \pname{} to explore the efficiency of ICCA chips. \pname{} also enables design space exploration of ICCA chip architecture.  
We demonstrate the capability of \pname{} using both an ICCA emulator and a simulator.

\vspace{0.2em}
\begin{acks}
\vspace{0.1em}
We thank the anonymous reviewers at MICRO'25 for their insightful feedback. 
We thank Michael Wang and Benjamin Reidys from the Systems Platform Research Group (Illinois PlatformX) at UIUC for proofreading our paper. This work was partially supported by the Hybrid Cloud and AI program at the IBM-Illinois Discovery Accelerator Institute (IIDAI), and NSF under the grants CAREER CNS-2144796, CCF-2107470, and CCF-1919044. 
\end{acks}

%% file: appendix.tex
\appendix

\newpage
\section{Artifact Appendix}

\def\mem{200\xspace}
\def\disk{20\xspace}
\def\hour{30\xspace}

\subsection{Abstract}

In this artifact, we provide the source code of \pname{}'s compilation, simulation, and evaluation framework. Then, we guide readers to explore how \pname{} improves the model serving performance on a variety of ICCA chips (i.e., Figure 17-24 in this paper).
To run this artifact, please use a Linux machine with at least \mem{} GB of main memory and at least \disk{} GB of disk space.

\subsection{Artifact Checklist (Meta-Information)}

{\small
\begin{itemize}[leftmargin=*]
  \item {\bf Algorithm:} Inductive tensor operator scheduling, cost-aware on-chip memory allocation, and ICCA chip design space exploration.
  \item {\bf Neural Network Models:} Llama2-13B, Gemma2-27B, OPT-30B, Llama2-70B, and DiT-XL. Their execution graphs are included in the repo.
  \item {\bf Run-time environment:} Ubuntu 20.04 or newer, Python 3.10.
  \item {\bf Metrics:} Execution time, hardware utilization.
  \item {\bf Output:} Trace files and result figures.
  \item {\bf Experiments:} Generate experiments using supplied scripts.
  \item {\bf How much main memory required (approximately):} \mem{} GB
  \item {\bf How much disk space required (approximately):} \disk{} GB
  \item {\bf How much time to prepare workflow (approximately):} 10 minutes
  \item {\bf How much time to complete experiments (approximately):} \hour{} hours on a machine with 64 CPU threads and 200 GB main memory.
  \item {\bf Publicly available:} Yes
  \item {\bf Archived (provide DOI):} 10.5281/zenodo.16541972
\end{itemize}
}

\subsection{Description}

\subsubsection{How to Access}

The source code can be downloaded from Zenodo at \url{https://doi.org/10.5281/zenodo.16541972}. For the latest version, you can access our GitHub repository: \url{https://github.com/platformxlab/elk.git}.

\subsubsection{Hardware Dependencies}

The \pname{} simulation and evaluation framework can run on any x86 machine with at least \mem{} GB of main memory and at least \disk{} GB of disk space.

\subsubsection{Software Dependencies}

The framework needs a Linux environment (preferably Ubuntu) with Python 3.10 installed.

\subsection{Installation}

\lstdefinestyle{BashStyle} {frame=tb,
  language=bash,
  aboveskip=3mm,
  belowskip=3mm,
  showstringspaces=false,
  columns=flexible,
  basicstyle={\footnotesize\ttfamily},
  numbers=none,
  numbersep=0pt,
  numberstyle=\tiny\color{gray},
  keywordstyle=\color{blue},
  commentstyle=\color{commentgreen},
  stringstyle=\color{mauve},
  breaklines=true,
  breakatwhitespace=true,
  tabsize=5,
  classoffset=0,
  keywordstyle=\color{blue},
}

\begin{enumerate}
    \item Start by downloading the \pname{} artifact from GitHub:
    \begin{lstlisting}[style=BashStyle,escapechar=~~,label=code:instrumentation]
    git clone https://github.com/platformxlab/elk.git
    cd elk
    \end{lstlisting}
    \item Please make sure all prerequisites are successfully installed:
    \begin{lstlisting}[style=BashStyle,escapechar=~~,label=code:instrumentation]
    sudo add-apt-repository ppa:deadsnakes/ppa
    sudo apt update
    sudo apt install python3.10 tmux -y
    curl -sS https://bootstrap.pypa.io/get-pip.py | python3.10
    python3.10 -m pip install -r requirements.txt
    \end{lstlisting}
\end{enumerate}

\vspace*{-0.3ex}
\subsection{Experiment Workflow}

To compile DL models into programs and obtain program execution traces from the \pname{} simulator, we provide a one-click script "\textstt{benchmark\_scripts/generate\_data\_from\_sim.py}" for you to launch all test cases in one place.
However, the script may take more than \hour{} hours to finish, so we recommend using \textstt{tmux}:
\begin{lstlisting}[style=BashStyle,escapechar=~~,label=code:instrumentation]
    tmux
\end{lstlisting}
Then, run the one-click script within the new \textstt{tmux} terminal:
\begin{lstlisting}[style=BashStyle,escapechar=~~,label=code:instrumentation]
    python3.10 benchmark_scripts/generate_data_from_sim.py
\end{lstlisting}
To return from the \textstt{tmux} terminal without pausing the script, press "Ctrl+B" and then press "D" on your keyboard.
To attach back to the original \textstt{tmux} terminal where the script is running, use:
\begin{lstlisting}[style=BashStyle,escapechar=~~,label=code:instrumentation]
    tmux attach -t 0
\end{lstlisting}
For more tips on using \textstt{tmux}, refer to \url{https://tmuxcheatsheet.com}.

\subsubsection{\textbf{Handle Errors}}
If the script encounters an error, the most common cause is that the artifact runs on too many CPU cores and overflows the main memory.
In such events, (1) go to "\textstt{launch.py}", (2) change the "\textstt{CORE\_REDUCE}" macro in line 22 to a larger value (e.g., \textstt{CORE\_REDUCE=8}), and (3) rerun the script:
\begin{lstlisting}[style=BashStyle,escapechar=~~,label=code:instrumentation]
    python3.10 benchmark_scripts/generate_data_from_sim.py
\end{lstlisting}
The script should automatically skip any completed test cases and resume from the failed one.

\subsection{Evaluation and Expected Results}

After the completion of all experiments, please run the following script to evaluate the results:
\begin{lstlisting}[style=BashStyle,escapechar=~~,label=code:instrumentation]
    ./run_artifact_eval_graph_gen.sh
\end{lstlisting}
This script gathers all data from the execution trace and draws all figures. To verify the results, one can compare the generated figures with those in the paper.

\subsection{Methodology}

Submission, reviewing and badging methodology:

\begin{itemize}
  \item \url{https://www.acm.org/publications/policies/artifact-review-and-badging-current}
  \item \url{https://cTuning.org/ae}
\end{itemize}

\vspace{3ex}